

\font\bigbold=cmbx12
\font\ninerm=cmr9      \font\eightrm=cmr8    \font\sixrm=cmr6
\font\fiverm=cmr5
\font\ninebf=cmbx9     \font\eightbf=cmbx8   \font\sixbf=cmbx6
\font\fivebf=cmbx5
\font\ninei=cmmi9      \skewchar\ninei='177  \font\eighti=cmmi8
\skewchar\eighti='177  \font\sixi=cmmi6      \skewchar\sixi='177
\font\fivei=cmmi5
\font\ninesy=cmsy9     \skewchar\ninesy='60  \font\eightsy=cmsy8
\skewchar\eightsy='60  \font\sixsy=cmsy6     \skewchar\sixsy='60
\font\fivesy=cmsy5     \font\nineit=cmti9    \font\eightit=cmti8
\font\ninesl=cmsl9     \font\eightsl=cmsl8
\font\ninett=cmtt9     \font\eighttt=cmtt8
\font\tenfrak=eufm10   \font\ninefrak=eufm9  \font\eightfrak=eufm8
\font\sevenfrak=eufm7  \font\fivefrak=eufm5
\font\tenbb=msbm10     \font\ninebb=msbm9    \font\eightbb=msbm8
\font\sevenbb=msbm7    \font\fivebb=msbm5
\font\tenssf=cmss10    \font\ninessf=cmss9   \font\eightssf=cmss8
\font\tensmc=cmcsc10

\newfam\bbfam   \textfont\bbfam=\tenbb \scriptfont\bbfam=\sevenbb
\scriptscriptfont\bbfam=\fivebb  \def\Bbb{\fam\bbfam}
\newfam\frakfam  \textfont\frakfam=\tenfrak \scriptfont\frakfam=%
\sevenfrak \scriptscriptfont\frakfam=\fivefrak  \def\frak{\fam\frakfam}
\newfam\ssffam  \textfont\ssffam=\tenssf \scriptfont\ssffam=\ninessf
\scriptscriptfont\ssffam=\eightssf  
\def\smc{\tensmc}

\def\eightpoint{\textfont0=\eightrm \scriptfont0=\sixrm
\scriptscriptfont0=\fiverm  \def\rm{\fam0\eightrm}%
\textfont1=\eighti \scriptfont1=\sixi \scriptscriptfont1=\fivei
\def\oldstyle{\fam1\eighti}\textfont2=\eightsy
\scriptfont2=\sixsy \scriptscriptfont2=\fivesy
\textfont\itfam=\eightit         \def\it{\fam\itfam\eightit}%
\textfont\slfam=\eightsl         \def\sl{\fam\slfam\eightsl}%
\textfont\ttfam=\eighttt         \def\tt{\fam\ttfam\eighttt}%
\textfont\frakfam=\eightfrak     \def\frak{\fam\frakfam\eightfrak}%
\textfont\bbfam=\eightbb         \def\Bbb{\fam\bbfam\eightbb}%
\textfont\bffam=\eightbf         \scriptfont\bffam=\sixbf
\scriptscriptfont\bffam=\fivebf  \def\bf{\fam\bffam\eightbf}%
\abovedisplayskip=9pt plus 2pt minus 6pt   \belowdisplayskip=%
\abovedisplayskip  \abovedisplayshortskip=0pt plus 2pt
\belowdisplayshortskip=5pt plus2pt minus 3pt  \smallskipamount=%
2pt plus 1pt minus 1pt  \medskipamount=4pt plus 2pt minus 2pt
\bigskipamount=9pt plus4pt minus 4pt  \setbox\strutbox=%
\hbox{\vrule height 7pt depth 2pt width 0pt}%
\normalbaselineskip=9pt \normalbaselines \rm}

\def\ninepoint{\textfont0=\ninerm \scriptfont0=\sixrm
\scriptscriptfont0=\fiverm  \def\rm{\fam0\ninerm}\textfont1=\ninei
\scriptfont1=\sixi \scriptscriptfont1=\fivei \def\oldstyle%
{\fam1\ninei}\textfont2=\ninesy \scriptfont2=\sixsy
\scriptscriptfont2=\fivesy
\textfont\itfam=\nineit          \def\it{\fam\itfam\nineit}%
\textfont\slfam=\ninesl          \def\sl{\fam\slfam\ninesl}%
\textfont\ttfam=\ninett          \def\tt{\fam\ttfam\ninett}%
\textfont\frakfam=\ninefrak      \def\frak{\fam\frakfam\ninefrak}%
\textfont\bbfam=\ninebb          \def\Bbb{\fam\bbfam\ninebb}%
\textfont\bffam=\ninebf          \scriptfont\bffam=\sixbf
\scriptscriptfont\bffam=\fivebf  \def\bf{\fam\bffam\ninebf}%
\abovedisplayskip=10pt plus 2pt minus 6pt \belowdisplayskip=%
\abovedisplayskip  \abovedisplayshortskip=0pt plus 2pt
\belowdisplayshortskip=5pt plus2pt minus 3pt  \smallskipamount=%
2pt plus 1pt minus 1pt  \medskipamount=4pt plus 2pt minus 2pt
\bigskipamount=10pt plus4pt minus 4pt  \setbox\strutbox=%
\hbox{\vrule height 7pt depth 2pt width 0pt}%
\normalbaselineskip=10pt \normalbaselines \rm}

\global\newcount\secno \global\secno=0 \global\newcount\meqno
\global\meqno=1 \global\newcount\appno \global\appno=0
\newwrite\eqmac \def\romappno{\ifcase\appno\or A\or B\or C\or D\or
E\or F\or G\or H\or I\or J\or K\or L\or M\or N\or O\or P\or Q\or
R\or S\or T\or U\or V\or W\or X\or Y\or Z\fi}
\def\eqn#1{ \ifnum\secno>0 \eqno(\the\secno.\the\meqno)
\xdef#1{\the\secno.\the\meqno} \else\ifnum\appno>0
\eqno({\rm\romappno}.\the\meqno)\xdef#1{{\rm\romappno}.\the\meqno}
\else \eqno(\the\meqno)\xdef#1{\the\meqno} \fi \fi
\global\advance\meqno by1 }

\global\newcount\refno \global\refno=1 \newwrite\reffile
\newwrite\refmac \newlinechar=`\^^J \def\ref#1#2%
{\the\refno\nref#1{#2}} \def\nref#1#2{\xdef#1{\the\refno}
\ifnum\refno=1\immediate\openout\reffile=refs.tmp\fi
\immediate\write\reffile{\noexpand\item{[\noexpand#1]\ }#2\noexpand%
\nobreak.} \immediate\write\refmac{\def\noexpand#1{\the\refno}}
\global\advance\refno by1} \def\semi{;\hfil\noexpand\break ^^J}
\def\nl{\hfil\noexpand\break ^^J} \def\refn#1#2{\nref#1{#2}}

\def\ann#1#2#3{{\it Ann.\ Phys.}\ {\bf {#1}} ({#2}) #3}

\def\plA#1#2#3{{\it Phys.\ Lett.}\ {\bf {#1}A} ({#2}) #3}

\newif\iftitlepage \titlepagetrue \newtoks\titlepagefoot
\titlepagefoot={\hfil} \newtoks\otherpagesfoot \otherpagesfoot=%
{\hfil\tenrm\folio\hfil} \footline={\iftitlepage\the\titlepagefoot%
\global\titlepagefalse \else\the\otherpagesfoot\fi}

\def\abstract#1{{\parindent=30pt\narrower\noindent\ninepoint\openup
2pt #1\par}}

\newcount\notenumber\notenumber=1 \def\note#1
{\unskip\footnote{$^{\the\notenumber}$} {\eightpoint\openup 1pt #1}
\global\advance\notenumber by 1}

\def\today{\ifcase\month\or January\or February\or March\or
April\or May\or June\or July\or August\or September\or October\or
November\or December\fi \space\number\day, \number\year}

\def\pagewidth#1{\hsize= #1}  \def\pageheight#1{\vsize= #1}
\def\hcorrection#1{\advance\hoffset by #1}
\def\vcorrection#1{\advance\voffset by #1}

\pageheight{23cm}
\pagewidth{15.7cm}
\hcorrection{-1mm}
\magnification= \magstep1
\parskip=5pt plus 1pt minus 1pt
\tolerance 8000
\def\bsk{\baselineskip= 14.5pt plus 1pt minus 1pt}
\bsk

\def\frac#1#2{{#1\over#2}}

\def\pmb#1{\setbox0=\hbox{$#1$} \kern-.025em\copy0\kern-\wd0
    \kern.05em\copy0\kern-\wd0 \kern-.025em\raise.0433em\box0 }

\def\ve{\vfill\eject}

\def\Z{{\Zed}}
\def\R{{\Real}\!}
\def\C{{\Complex}}
\def\({\left(}
\def\){\right)}

\def\R{{\Bbb R}}    \def\Z{{\Bbb Z}}  \def\C{{\Bbb C}}

\def\figone{twoline.eps}
\def\figtwo{twointav.eps}
\def\figthree{twospectline.eps}

\input epsf

\let\omitpictures=N


{

\refn\RS
{M. Reed, B. Simon,
\lq\lq Methods of Modern Mathematical Physics\rq\rq , 
{\sl Vol.II}, Academic Press, New York, 1980}

\refn\AGHH
{S. Albeverio, F. Gesztesy, R. H{\o}egh-Krohn and H. Holden,
\lq\lq Solvable Models in Quantum Mechanics\rq\rq,
Springer, New York, 1988}

\refn\Seba
{P. \v{S}eba,
{\it Czech. J. Phys.} {\bf 36} (1986) 667
}
\refn\TFC
{I. Tsutsui, T. F\"{u}l\"{o}p and T. Cheon,
 {\it J. Phys. Soc. Jpn.} {\bf 69} (2000) 3473}

\refn\CFT
{T. Cheon, T. F\"{u}l\"{o}p and I. Tsutsui,
\ann{294}{2001}{1}}

\refn\TUIT
{T. Uchino and I. Tsutsui,
{\sl Supersymmetric Quantum Mechanics with a Point Singularity},
quant-ph/0210084}

\refn\Kobe
{T. Nagasawa, M. Sakamoto and K. Takenaga,
{\sl Supersymmetry in Quantum Mechanics with Point Interactions},
hep-th/0212192}

\refn\CKS
{F. Cooper, A. Khare and U. Sukhatme, 
 {\it Phys. Rep.} {\bf 251} (1995) 267}

\refn\Junker
{G. Junker,
\lq\lq Supersymmetric Methods in Quantum
and Statistical
Physics\rq\rq,
Springer,
Berlin, 1996}

\refn\CKStwo
{F. Cooper, A. Khare and U. Sukhatme, 
\lq\lq Supersymmetry in Quantum
Mechanics\rq\rq,
World Scientific, Singapole, 2001}


\refn\Sukumar
{C. V. Sukumar,
{\it J. Phys. A: Math. Gen.} {\bf 18} (1985) L57}

\refn\GLR
{J. Goldstein and C. Lebiedzik and R. W. Robinett,
{\it Am. J. Phys.} {\bf 62} (1994) 612}

\refn\VA
{S. De Vincenzo and V. Alonso,
\plA{298}{2002}{98}}

\refn\Witone
{E. Witten,
{\it Nucl. Phys.} {\bf B 185} (1981) 513}

\refn\Wittwo
{E. Witten,
{\it Nucl. Phys.} {\bf B202} (1982) 253}

\refn\AG
{N.I. Akhiezer and I.M. Glazman,
\lq\lq Theory of Linear Operators in Hilbert Space\rq\rq,
{\sl Vol.II},
Pitman Advanced Publishing Program, Boston, 1981}

\refn\FT
{T. F\"{u}l\"{o}p and I. Tsutsui,
\plA{264}{2000}{366}}

\refn\Boya
{L. J. Boya,
{\it Eur. J. Phys.} {\bf 9} (1988) 139}

\refn\TCF
{I. Tsutsui, T. Cheon and T. F\"{u}l\"{o}p,
{\it J. Phys. A: Math. Gen.} {\bf 36} (2003) 275}

}

\pageheight{23cm}
\pagewidth{15.7cm}
\hcorrection{-1mm}
\magnification= \magstep1
\parskip=5pt plus 1pt minus 1pt
\tolerance 8000
\baselineskip= 15pt plus 1pt minus 1pt



\pageheight{23cm}
\pagewidth{15.7cm}
\hcorrection{0mm}
\magnification= \magstep1
\def\bsk{%
\baselineskip= 16.8pt plus 1pt minus 1pt}
\parskip=5pt plus 1pt minus 1pt
\tolerance 6000




\hfill 
{KEK Preprint 2002-133}
\vskip -4pt 
\hfill \phantom{quant-ph/0207xxx}

\vskip 42pt

{\baselineskip=18pt

\centerline{\bigbold
Supersymmetric Quantum Mechanics}
\centerline{\bigbold
under Point Singularities}

\vskip 30pt

\centerline{\smc
Takashi Uchino%
\quad
{\rm and}
\quad
Izumi Tsutsui\footnote{${}^\dagger$}
{\eightpoint email:\quad izumi.tsutsui@kek.jp}
}

\vskip 7pt

{
\baselineskip=13pt
\centerline{\it
Institute of Particle and Nuclear Studies}
\centerline{\it
High Energy Accelerator Research Organization (KEK)}
\centerline{\it Tsukuba 305-0801}
\centerline{\it Japan}
}

\vskip 80pt

\abstract{%
{\bf Abstract.}\quad
We provide a systematic study on the possibility of supersymmetry (SUSY) for one dimensional quantum mechanical systems consisting of a pair of lines $\R$ or intervals  $[-l, l]$ each having a point singularity. We consider the most general singularities and walls (boundaries) at $x = \pm l$ admitted quantum mechanically, using a $U(2)$ family of parameters to specify one singularity and similarly a $U(1)$ family of parameters to specify one wall.   With these parameter freedoms, we find that for a certain subfamily the line systems acquire an $N = 1$ SUSY which can be enhanced to $N = 4$ if the parameters are further tuned, and that these SUSY are generically broken except for a special case.  The interval systems, on the other hand, can accommodate $N = 2$ or $N = 4$ SUSY,  broken or unbroken, and exhibit a rich variety of (degenerate) spectra. Our SUSY systems include the familiar SUSY systems with the Dirac $\delta(x)$-potential, and hence are extensions of the known SUSY quantum mechanics to those with general point singularities and walls.  The self-adjointness of the supercharge in relation to the self-adjointness of the Hamiltonian is also discussed.}

\vskip 10pt
%
%
%
%
}
\ve


\pageheight{23cm}
\pagewidth{15.7cm}
\hcorrection{-1mm}
\magnification= \magstep1
\def\bsk{%
\baselineskip= 15pt plus 1pt minus 1pt}
\parskip=5pt plus 1pt minus 1pt
\tolerance 8000
\bsk


\secno=1 \meqno=1

\bigskip
\noindent{\bf 1. Introduction}
\medskip

A point singularity (or interaction) appears 
in various different 
contexts in physics.  It may, for instance, appear as a point defect or a junction of two
layers of materials, or may be considered as a localized limit of a finite range
potential in general.  A point singularity is usually modelled by the Dirac
$\delta({\bf x})$-potential, which offers exact solutions to
a number of problems of interest both classically and quantum mechanically.
However, 
in quantum mechanics a point singularity is far from unique --- 
in one dimension, for example,  the Dirac
$\delta(x)$ is just one of the $U(2)$ family of point
singularities allowed quantum mechanically
[\RS, \AGHH, \Seba].  
In fact, recent investigations have shown
that these point
singularities can give rise to unexpectedly interesting phenomena which are
not available under the Dirac
$\delta(x)$-potential.  These include duality in spectra,
anholonomy (Berry phase) and scale anomaly [\TFC, \CFT], which normally occur in more
complicated systems or quantum field theory.  

The first of these, the
duality, implies that the spectra of two distinct point singularities may coincide
if they are related under some discrete transformations and, in particular,
become completely degenerate if the point singularity is self-dual, {\it
i.e.},  invariant under the discrete transformations.  The presence of degeneracy, and
also the graded structure which is naturally equipped with the system,  alluded us to
examine the possibility of supersymmetry (SUSY) with self-dual point singularities.
This has indeed
been confirmed in our previous paper [\TUIT], where we have found 
a number of novel  $N = 1$ and $N = 2$ SUSY
systems on a line $\R$ or interval $[-l, l]$ with the family of $U(2)$ point singularities
(under the walls at $x = \pm l$ 
allowing for the general boundary condition for the interval
case).  More recently, SUSY on a circle with two point singularities has also been studied in
[\Kobe].

Meanwhile, SUSY quantum mechanics (and its extensions) has been studied
intensively  over the years, 
initially to
provide SUSY breaking mechanisms in field theory and lately to establish
schemes to accommodate known solvable
models or generate novel ones (see, {\it e.g.}, [\CKS, \Junker,
\CKStwo]). However, for some reason the investigation of SUSY quantum mechanics under
point singularities has evaded from the studies, and we know little
about it except that a suitable pair\note{%
There are works [\Sukumar, \GLR, \VA] on SUSY quantum mechanics for a pair of
interval systems one of which has the Dirac
$\delta(x)$-potential while the other  has a \lq partner potential\rq{} with a finite
support.  In contrast, here we will consider a pair of line/interval systems
possessing singularities without such potentials.} 
of the Dirac
$\delta(x)$-potentials can be made into a Witten model [\Witone,
\Wittwo] and realizes an $N = 2$ SUSY. The aim of this paper is to present a systematic
study of SUSY under point singularities, and thereby report that point
singularities admit a variety of novel SUSY systems including the known one.  The systems
we consider are those consisting of a pair of lines
$\R$ or intervals
$[-l, l]$ each having a point singularity, where the pair provides a 
graded
structure as the known Dirac
$\delta(x)$ system has.  The two point singularities
can in general be different and hence our total family of singularities
are given by 
$U(2)
\times U(2)$.  We find that the line
systems with point singularities belonging to a certain subfamily 
generically possess an $N
= 1$ SUSY, broken and unbroken, and that the SUSY can be promoted to
$N = 4$ for a further restricted parameter subfamily.  Similarly, for the interval systems, 
in addition to
the general $U(2) \times U(2)$ point singularities
we take account of the most general boundary conditions for the two sets of walls at 
$x = \pm l$ represented by
$[U(1)]^4$.  We then find that an 
$N = 2$ or
$N = 4$ (broken or unbroken) SUSY appears for a certain subfamily of the combined
parameter family characterizing the point
singularities and the walls of the system.  All of these SUSY systems are classified into
a number of different types, and their SUSY and spectral properties are summarized in
Appendix B.

The plan of the paper is as follows.  In section 2, we discuss the line systems, where we
provide our criterion for SUSY and thereby find SUSY systems ((A1) -- (B2)).  Based on the
result of this section, the interval systems are then studied in section 3, where we
find quite a few distinct SUSY systems ((a1) -- (d6)).  Section 4 is devoted to 
the question of the self-adjointness of the supercharge.  Our conclusion
and discussions are given in section 5.  Appendix A contains computations to
supplement our argument in section 4, and Appendix B furnishes the summary table 
for the various SUSY systems mentioned above.

\ve
\centerline{{\bf 2. Two lines with point singularity }}
\medskip

\secno = 2 \meqno=1

In this section, we investigate the possibility of
supersymmetry in a quantum system consisting of 
two lines each possessing a
singularity at 
$x = 0$.  The Hilbert space of our system is thus given by
${\cal H} =
L^2(\R\backslash\{0\})\oplus
L^2(\R\backslash\{0\}) \simeq L^2(\R\backslash\{0\})\otimes\C^2$.  Except for
the singular point which is now removed on each of the lines, the system is
assumed to be free and hence its Hamiltonian reads
$H = -\frac{\hbar^2}{2m}\frac{d^2}{dx^2}\otimes I_2$, where $I_2$
denotes the $2\times 2$ unit matrix.  
If we split the space in two at the singularity and thereby regard
$L^2(\R\backslash\{0\})
\simeq L^2(\R^+)\otimes\C^2$, we can double the graded structure and identify
${\cal H}$ with
$L^2(\R^+)\otimes\C^4$ (see Fig.1).
Let $\psi_1(x)$ and $\psi_2(x)$ be the wave functions on the two lines,
respectively, representing a state in the Hilbert space.  According to the
above identification, the state may equally be represented by
$\Psi_i(x) = (\psi_i^+(x), \psi_i^-(x))^T$ for $i =1$, 2, where we have
defined
$\psi_i^+(x) = \psi_i(x)$ for $x > 0$ and $\psi_i^-(-x) = \psi_i(x)$ for
$x < 0$.  Combining these, we can express a state in the Hilbert space
${\cal H}$ by the four-components wave
function,
$$
\Psi(x) = 
	\left( \matrix{
		\Psi_1(x) \cr
		\Psi_2(x)
	} \right),
\qquad x > 0,
\eqn\fun
$$
on which our Hamiltonian takes the form,
$$
H = -\frac{\hbar^2}{2m}\frac{d^2}{dx^2}\otimes I,
\qquad 
\hbox{where} \quad
I =
	\left( \matrix{ 
		I_2 & { 0} \cr
		{ 0} & I_2
	} \right).
\eqn\hamihami
$$

The first question we need to address is to find a proper domain on which
the Hamiltonian (\hamihami) becomes self-adjoint.  This is 
ensured by requiring the probability current 
$
j^\pm_i(x) =
{{i\hbar}\over{2m}}[(\psi^\pm_i)^\dagger
(\psi^\pm_i)^\prime - (\psi^\pm_i)^{\prime\dagger} \psi^\pm_i](x)
$ 
be continuous over the two
lines (the dash denotes the derivative $\psi' = {d\over{dx}}\psi$).  If the two
singular points, $x = +0$ on line 1 and
$x = +0$ on line 2, were connected at one point, then the continuity condition at the
point would be (for brevity we hereafter denote $x = 0$ for $x = +0$)
$$
0 = \sum_{a = \pm} j^a_1(0) + \sum_{a = \pm} j^a_2(0) =
{{i\hbar}\over{2m}}(\Psi^\dagger
\Psi^\prime - (\Psi^\prime)^\dagger\Psi)( 0).
\eqn\cons
$$
Introducing an arbitrary real constant $L_0 \neq 0$
with dimension of length, 
we find that the condition (\cons) is equal to $|\Psi(0) +
iL_0\Psi^\prime(0)| = |\Psi(0) - iL_0\Psi^\prime(0)|$.   
This in turn can be written as [\AG, \FT]
$$
(U-I)\Psi(0) + iL_0(U+I)\Psi^\prime(0) = 0,
\eqn\bouzu
$$
with a matrix $U \in U(4)$, or equivalently, 
$$
\Psi^{(-)}(0) = U\, \Psi^{(+)}(0),
\eqn\zoom
$$ 
in terms of $\Psi^{(\pm)} = \Psi \pm iL_0\Psi^\prime$.
The matrix $U$, which is called \lq
characteristic matrix\rq{} since it characterizes the nature of singularity,
specifies a self-adjoint domain 
${\cal D}_U(H) \subset {\cal H}$ of the Hamiltonian (\hamihami) by means of
the boundary conditions at the two singular points
connected.  Of course, in our system the two points are disconnected, and
hence the actual continuity condition is 
$$
\sum_{a = \pm} j^a_1(0) = 0 = \sum_{a = \pm} j^a_2(0).
\eqn\contcond
$$
Correspondingly,
our unitary matrix $U$ must be specialized to $U \in U(2) \times U(2)
\subset U(4)$, namely, 
$$
U =
	\left( \matrix{ 
		U_1 & { 0} \cr
		{ 0} & U_2
	} \right),
\qquad U_1, \,\, U_2 \in U(2).
\eqn\bloblo
$$
Despite the specialization, we prefer to work in the four-components
description which is more convenient on account of the fact that all of the
components of $\Psi(x)$ may be interchanged under SUSY transformations.

\topinsert
\epsfxsize 6.5cm
\ifx\omitpictures N   \centerline{\epsfbox {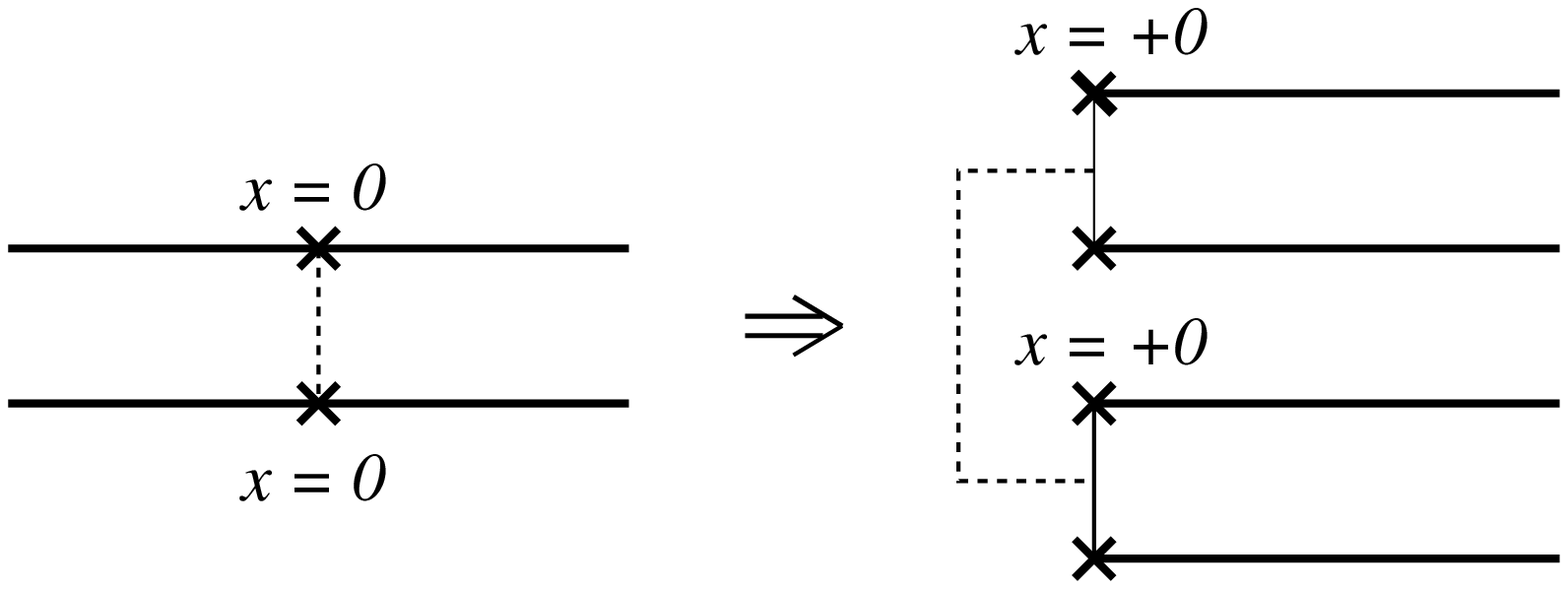}}  \fi
\abstract{{\bf Figure 1.}~A system of a pair of two 
lines each having a singularity at $x = 0$ may be identified with two systems
of a pair of two half lines where the probability flow is allowed to pass 
between the two systems through $x = 0$. }
\bigskip
\endinsert

Our next task is to seek SUSY possible under the
Hamiltonian (\hamihami) with domain specified by (\zoom).  
We suppose that
the supercharge
$Q$ be a self-adjoint operator (which will be examined later) 
and of the form, 
$$
Q = -i\lambda\frac{d}{dx}\otimes\Gamma + \mu\otimes\Omega,
\eqn\supsup
$$
where we set $\lambda = \hbar/(2\sqrt{m})$ and $\mu$ is a real constant.  
$\Gamma$ and $\Omega$ are Hermitian $4\times4$
matrices and assumed to satisfy the conditions,
$$
\Gamma^2 = I, \qquad \Omega^2 = I, 
\qquad \{\Gamma, \,\Omega\} = 0,
\eqn\coec
$$
which lead to
$2Q^2 = H + \mu^2$ (more precisely, $\mu^2$ should be written as 
$\mu^2\cdot {\rm id}_{\cal H}$ where  ${\rm id}_{\cal H}$ denotes the identity
operator in the Hilbert space ${\cal H}$) with the Hamiltonian $H$ in (\hamihami).  The
extra term
$\mu^2$ can then be absorbed into $H$ by redefining $H + \mu^2 \rightarrow H$, {\it
i.e.}, by  the constant energy shift by $\mu^2$, to realize the standard SUSY relation
$$
2Q^2 = H.
\eqn\ancc
$$
Note that this redefinition does not alter the   
domain 
${\cal D}_U(H)$ because it does not affect our argument of the 
probability conservation.

Our aim now is to find a SUSY invariant pair $(U, Q)$ in the sense
that
the SUSY transformation generated by the supercharge $Q$ in (\supsup)
preserves the domain of each energy eigenstate $\Phi$
with\note{%
The equation does not necessarily imply degeneracy in the energy level, because
some of the components in the vector eigenstate $\Phi$ may vanish.} 
$$
H \,\Phi(x) = E\, \Phi(x),
\eqn\eeigenm
$$
that is, we say that the pair $(U, Q)$ is \lq SUSY invariant\rq{}
$$
\hbox{if} \quad 
\Phi \in {\cal D}_U(H)
\qquad
\hbox{then}
\quad Q\,\Phi \in {\cal D}_U(H),
\eqn\invpair
$$ 
for
$\Phi$ satisfying (\eeigenm).
Our task to find such a pair may considerably be simplified if we recall
that any
$U(4)$ matrix
$U$ can be decomposed as
$U=V^{-1}DV$ with an $SU(4)$ matrix $V$ and a diagonal matrix,
$$
D = \hbox{diag}\,(e^{i\theta_1}, e^{i\theta_2}, e^{i\theta_3},
e^{i\theta_4}), \qquad \theta_k \in [0, 2\pi), \quad k = 1, \ldots, 4.
\eqn\monmon
$$
With this decomposition, we see from the boundary conditions (\bouzu) that
if
$\Psi(x)
\in {\cal D}_U(H)$ then $W\Psi(x) \in {\cal D}_{WUW^{-1}}(H)$ for any $W \in
SU(4)$. Hence, if a pair $(U, Q)$ satisfies the SUSY invariant condition
(\invpair), then
$(WUW^{-1}, WQW^{-1})$ also satisfies the condition (note
that 
$WQW^{-1}$ is again of the form (\supsup)). 
Choosing in particular $W
= V$, we can obtain a pair $(D, VQV^{-1})$.  This implies that, if 
a pair $(D, Q)$ is a solution, so is $(U, V^{-1}QV)$.  Thus our aim is
achieved if we obtain a solution for the diagonal case $(D, Q)$ first, and
then transform it to $(U, V^{-1}QV)$ with 
$$
V =
	\left( \matrix{ 
		V_1 & { 0} \cr
		{ 0} & V_2
	} \right),
\qquad V_1, \,\, V_2 \in SU(2),
\eqn\vbloblo
$$
for which $U=V^{-1}DV$ has the block diagonal form (\bloblo).
We also note that the order of the factors $e^{i\theta_i}$ in $D$ in
(\monmon) is unimportant, because any $D$ can be put into
$D = S^{-1}\bar D S$ with some exchange matrix $S$ so that 
$\bar D$ has a desired order of the factors.  
The exchange matrix $S$ may be absorbed
into $V$ by redefining $V \rightarrow S^{-1}V$, if $S$ is
block diagonal as in $V$ in (\vbloblo).  If not, we need to keep $S$ as an
additional element in the decomposition of $U$ when we use the ordered $\bar D$. 
We record, however, only two cases which we use later,
namely, 
$S = X$ or
$Y$ where
$$
X = 
	\left( \matrix{
		0 & 0 & 1 & 0 \cr
		0 & 0 & 0 & 1 \cr
		1 & 0 & 0 & 0 \cr
		0 & 1 & 0 & 0
	} \right),
\qquad
Y = 
	\left( \matrix{
		1 & 0 & 0 & 0 \cr
		0 & 0 & 1 & 0 \cr
		0 & 1 & 0 & 0 \cr
		0 & 0 & 0 & 1
	} \right).
\eqn\esmtrx
$$
In short, the characteristic matrix $U$ may be decomposed as
$$
U = V^{-1} D V,
\qquad D = S^{-1} \bar D S,
\eqn\udecomp
$$
with
$\bar D$ having diagonal factors in a desired order, using an appropriate 
exchange matrix $S$ which
may (or may not) be absorbed in $V$ depending on the order one wants.
When we use $\bar D$ with $S$ and obtain a solution for the SUSY pair
$(\bar D, Q)$, we find the general solution by transforming it to
$(U, V^{-1}S^{-1}QSV)$ as before.

We now consider the SUSY transformation generated by the supercharge $Q$
in (\supsup) for states obeying the boundary condition (\bouzu) for
diagonal 
$U = D$.  Under the SUSY transformation, the eigenstate
$\Phi$ and its derivative are transformed into
$$
\eqalign{
	(Q\Phi)(x) 
&=
	-i\lambda\Gamma\Phi^\prime(x) + \mu\Omega\Phi(x), \cr	
	(Q\Phi^\prime)(x) 
&= 
	-i\lambda\Gamma\Phi^{\prime\prime}(x) + \mu\Omega\Phi^\prime(x)
	= iE\lambda\Gamma\Phi(x) + \mu\Omega\Phi^\prime(x).
}
\eqn\hirame
$$
If the transformed state
$Q\Phi$ is to satisfy the original boundary condition (\bouzu), we need
$$
\eqalign{
	&(D - I) \Gamma \bigl(\Phi^{(+)}(0) - \Phi^{(-)}(0)\bigr) 
	- 2 \mu L_0\bigl(D\Omega\Phi^{(+)}(0) - \Omega\Phi^{(-)}(0)\bigr)
\cr
	&\qquad\qquad\qquad\qquad\qquad + E L_0^2 (D + I) \Gamma \bigl(\Phi^{(+)}(0) +
\Phi^{(-)}(0)\bigr) = 0. }
\eqn\sawara
$$
Using (\zoom), the condition (\sawara) becomes
$$
\bigl(\lambda(D-I)\Gamma(D-I) + 2\mu L_0[D, \Omega] -
EL_0^2(D+I)\Gamma(D+I)\bigr)\Phi^{(+)}(0)= 0.
\eqn\sptc
$$
At this point it is important to recognize that the original condition (\zoom)
provides relations among the components between
$\Psi^{(+)}(0)$ and
$\Psi^{(-)}(0)$, but not among those within $\Psi^{(+)}(0)$ or $\Psi^{(-)}(0)$. 
It follows that, if the condition (\sptc) is identical to (\zoom), then 
the coefficient matrix for $\Phi^{(+)}(0)$ in (\sptc) must vanish. 
Furthermore, 
since the original condition (\bouzu) is energy independent, we require that
the equality (\sptc) holds independently of $E$.  
Thus the condition (\sptc) actually
implies
$$
(D + I) \Gamma (D + I) = 0 ,
\eqn\hamati
$$
and
$$
\lambda (D - I) \Gamma (D - I) + 2\mu L_0[D + I, \Omega] = 0,
\eqn\simaazi
$$
where for
our later convenience we have replaced $D$ with  $D+I$ in the second term.

From (\hamati) one immediately sees that at least one element of the diagonal
matrix
$D$ must be $-1$ (otherwise $D+I$ has an inverse and hence we obtain $\Gamma =
0$ in contradiction to (\coec)).  Let $n$ be the number of $e^{i\theta_k} = -1$
elements among the four in (\monmon).  With the help of the exchange
matrix $S$, we may chose the ordered diagonal 
$\bar D$ in (\udecomp) such that those $-1$ elements 
are arranged in the lower right corner, {\it i.e.,} 
$e^{i\theta_{4 - n + 1}} = \cdots = e^{i\theta_4} = -1$.  
Under this arrangement,
the submatrix of $\bar D + I$ given by its upper left
$(4 - n) \times (4-n)$ block has an inverse, and hence one observes in (\hamati) that all
the elements in the corresponding block in
$\Gamma$ must vanish.  On the other hand, since the submatrix of 
$\bar D + I$
given by its lower right $n \times n$ block vanishes identically, one finds from
(\simaazi) that the elements in the corresponding block in $\Gamma$ vanish, too. 
Combining these, one learns that
$\Gamma$ has nonvanishing elements 
only in the blocks other than these two.  Since
such a $\Gamma$ has $\det\Gamma \ne
0$
(which is required for $\Gamma^2 = I$ in (\coec)) only if
$n = 2$, a SUSY invariant pair can be found only
for the case,
$$
\bar D =
	\left( \matrix{ 
		T & { 0} \cr
		{ 0} & -{I_2}
	} \right),
\qquad 
T =
	\left( \matrix{ 
		e^{i\theta_1} & 0 \cr
		0 & e^{i\theta_2}
	} \right),
\qquad \theta_1, \, \theta_2 \ne \pi.
\eqn\dbloblo
$$
Here, $\Gamma$ takes the form,
$$
\Gamma =
	\left( \matrix{ 
		{ 0} & A \cr
		 A^\dagger & { 0}
	} \right),
\eqn\no
$$
where the $2 \times 2$ matrix $A$ is seen to be unitary, $A \in U(2)$, to
ensure the condition
$\Gamma^2 = I$. 

{}For $\mu \ne 0$ we further need to determine $\Omega$ and for this we first set
$$
\Omega =
	\left( \matrix{ 
		B & C \cr
		 C^\dagger & F
	} \right),
\qquad B = B^\dagger, \quad F = F^\dagger.
\eqn\megoti
$$
Then (\simaazi) requires that 
$$
[T, \,B] = 0, \qquad
C = iK A,
\eqn\no
$$
where we have used the real diagonal matrix,
$$
K = {{\lambda}\over{i\mu L_0}}(T+I_2)^{-1}(T-I_2)
= 	{\lambda\over{\mu}}
\left( \matrix{ 
		1/L(\theta_1) & 0 \cr
		 0 & 1/L(\theta_2)
	} \right),
\eqn\amaterasu
$$
which is given in terms of the two scale parameters [\CFT] defined by
$$
L(\theta_i) = L_0\cot{{\theta_i}\over 2}, \qquad i = 1, \,\, 2,
\eqn\spara
$$
which are nonvanishing $L(\theta_i) \ne 0$ because $\theta_i \ne \pi$
(see (\dbloblo)).
On the other hand, $\{\Gamma, \,\Omega\} = 0$ in (\coec) is ensured if
$$
{}F = -A^\dagger B A.
\eqn\no
$$
{}For the remaining condition $\Omega^2 = I$ in (\coec) to hold, in addition
to what we already have, we need only 
$$
B^2 = {I_2} - K^2.
\eqn\no
$$

We therefore arrive at the general solution of SUSY invariant systems, that is, 
a pair
$(D = S\bar D S^{-1}, Q)$ is SUSY invariant if
$\bar D$ has the form (\dbloblo) and $Q = q(A, \mu; K)$ with
$$
q(A, \mu; K) = -i\lambda\frac{d}{dx}\otimes	\left( \matrix{ 
		{ 0} & A \cr
		 A^\dagger & { 0}
	} \right) + \mu\otimes\left( \matrix{ 
		B & iK A \cr
		 -i A^\dagger K & -A^\dagger B A
	} \right),
\eqn\izanagi
$$
where $A \in U(2)$, ${K}$ is given by (\amaterasu) and $B$ is subject to
the conditions,
$$
\qquad B = B^\dagger, 
\qquad B^2 = {I_2} - {K}^2,
\qquad [T, \,B] = 0,
\eqn\izanami
$$
which are met by
$$
B = b_1 \sigma_+ + b_2 \sigma_-, 
\qquad \sigma_\pm = {{{I_2} \pm \sigma_3}\over 2}, 
\qquad
b_i^2 = 1 - \left[{{\lambda}\over{\mu L(\theta_i)}}\right]^2,
\eqn\yamatotakeru
$$
where $\sigma_i$, $i = 1$, 2, 3, are the Pauli matrices.
Besides, if $\theta_1 = \theta = \theta_2$, we have the additional solution,
$$
B = \sum_{i = 1}^3 b_i \sigma_i,
\qquad
\sum_{i = 1}^3 b_i^2 = 
1 - \left[{{\lambda}\over{\mu L(\theta)}}\right]^2.
\eqn\homudawake
$$
One can decouple the freedom of $A$ in $\Gamma$ and $\Omega$ by casting them into
$$
\Gamma = \Sigma^{-1}
	\left( \matrix{ 
		{ 0} & {I_2} \cr
		 {I_2} & { 0}
	} \right) \Sigma = \Sigma^{-1}({I_2} \otimes \sigma_1) \Sigma,
\eqn\fmgm
$$
and
$$
\Omega = \Sigma^{-1}
	\left( \matrix{ 
		B & i{K}  \cr
		 -i {K} & - B 
	} \right) \Sigma
= \Sigma^{-1}({K} \otimes \sigma_2 + B \otimes \sigma_3) \Sigma,
\eqn\no
$$
by introducing
$$
\Sigma =
	\left( \matrix{ 
		{I_2} & { 0}  \cr
		 { 0} & A 
	} \right).
\eqn\konahanasakuya
$$
In fact, $\Sigma$ is a special type of the unitary matrix $V$
in  (\vbloblo) which
leaves
$\bar D$ in (\dbloblo) invariant, 
$\bar D \rightarrow \Sigma^{-1} \bar D \Sigma = \bar D$, and this invariance
provides the freedom
$A$ in the choice of the supercharge $Q$.

Given a $D$ (and hence a ${K}$), there are at most four \lq independent\rq{}
supercharges among those 
$Q = q(A, \mu; {K})$ 
in (\izanagi) for
$A \in U(2)$, and a set of independent supercharges may be
furnished by  
$$
Q_k = q(i\sigma_k, \mu; {K}), \quad 
k = 1, 2, 3, \qquad 
Q_4 = q({I_2}, \mu;
{K}).
\eqn\otohimesama
$$
Because of the $\mu$-term in the supercharges, however, 
the standard orthogonal SUSY algebra,
$$
\{Q_i, \, Q_j\} = H \, \delta_{ij},
\eqn\ototachibana
$$
for all $i, \, j = 1, \ldots, 4$, cannot be realized unless $B = 0$, that is,
$$
\theta_1 = \theta = \theta_2 \quad\hbox{or} \quad 
\theta_1 = \theta = 2\pi - \theta_2, \qquad 
\mu^2 = \left[{{\lambda}\over{L(\theta)}}\right]^2.
\eqn\hondawake
$$
Thus, if we say that a system has an $N = n$ SUSY if it has 
$n$ supercharges satisfying (\ototachibana), then we see that 
our system can possess an $N = 4$ SUSY if the two angles $\theta_1$ and
$\theta_2$ in the characteristic matrix $U$ are related by (\hondawake) and further 
if we specify the parameter $\mu$ in the supercharge $Q$ by (\hondawake).  
Otherwise,  the system possesses only an $N = 1$ SUSY.  
Note that the set of supercharges satisfying the orthogonal relation
(\ototachibana) is not unique, since it can always be transformed by 
$Q_k \rightarrow
\Sigma^{-1} Q_k \Sigma$ with
$\Sigma$ in the form (\konahanasakuya) leaving the relation intact.

If $\mu = 0$, on the other hand,
one finds from (\hamati) and (\simaazi) that
$T = {I_2}$. 
This suggests that the SUSY system with a point singularity possessing
the standard SUSY algebra without
the
$\mu$-term in (\ototachibana) is basically unique --- a SUSY is
realized only if the characteristic matrix
$U$ has the diagonal part 
$D =
\hbox{diag}(1,1,-1,-1)$ modulo the possible exchanges of the $\pm 1$ elements.  
Systems on a circle with different combinations of these special types of 
point singularities
have been studied in detail in [\Kobe].

For illustration, we mention a simple but generic $N = 1$ case obtained by the 
diagonal 
$U = \bar D$ in (\dbloblo).  The boundary condition (\bouzu) then reads
$$
\psi^+_1(0) + L(\theta_1){\psi^+_1}^\prime(0) =0, 
\qquad
\psi^-_1(0) + L(\theta_2){\psi^-_1}^\prime(0) =0,
\qquad
\psi^+_2(0)=\psi^-_2(0)=0.
\eqn\no
$$
Apart from the travelling wave eigenstates which are four-fold degenerate,
one finds the two bound states,
$$
\Phi^{(1)}(x) = 
	\left( \matrix{
		e^{-x/L(\theta_1)} \cr
 0  \cr
	0  \cr
 0  \cr
	} \right),
\qquad
\Phi^{(2)}(x) = 
	\left( \matrix{
	0 \cr
 e^{-x/L(\theta_2)}  \cr
	0  \cr
 0  \cr
	} \right),
\eqn\exaboun
$$
which are allowed for
$L(\theta_1)>0$ and $L(\theta_2)>0$, respectively.

To sum up, we have found that a system of two lines whose singularity is characterized
by $U$ admits an $N = 1$ SUSY if the ordered diagonal matrix $\bar D$ 
in the decomposition (\udecomp) has the form (\dbloblo).  The supercharge
is provided by
$$
Q = V^{-1}S^{-1}q(A, \mu;{K})SV,
\eqn\gensc
$$
with $q(A, \mu; {K})$ given by (\izanagi).
The SUSY can be enhanced to $N = 4$ if $B = 0$, that is, if the conditions (\hondawake)
are fulfilled. These SUSY systems
exhibit distinct features depending on the choice
of the matrix $B$ and the angles $\theta_1$ and $\theta_2$.  In fact, the parameter
dependence of the features can be seen just by observing the example mentioned above,
because the energy spectrum depends only on the spectral parameters on account of the
conjugations on $U$ which preserve the spectrum.  We
classify the SUSY systems into the four types:
$$
\eqalign{
&{\rm (A1)} \qquad B\neq 0, \quad \theta_1\neq\theta_2,
\cr &{\rm (A2)} \qquad B\neq 0, \quad \theta_1=\theta_2,
\cr &{\rm (B1)} \qquad B=0, \quad L(\theta_1)>0
\,\,\,\, \hbox{or} \,\,\,\, L(\theta_2)>0, \cr 
&{\rm (B2)} \qquad B=0, \quad L(\theta_1)\le 0
\,\,\,\, \hbox{and} \,\,\,\, L(\theta_2)\le 0. 
}
$$
In type (A1) and (A2) systems, the bound states 
(\exaboun) are mapped into themselves under the SUSY transformation
generated by the supercharge (\izanagi) with the diagonal $B$ given in (\yamatotakeru).  
For type (A2) where the two angles $\theta_1$ and
$\theta_2$ coincide, one may also choose (\homudawake) with off-diagonal elements for $B$
so that the SUSY transformation induces the exchange
between $\Phi^{(1)}$ and $\Phi^{(2)}$.   In both of these types for which  $B\neq
0$, therefore, the SUSY is broken.   In contrast, the ground state in type (B1), which
admits at least one bound state, is annihilated under the SUSY transformation, and
hence the SUSY is good ({\it i.e.}, unbroken).  Type (B2) possesses only degenerate 
positive energy states which
are related by the SUSY
transformation, and the SUSY is broken.  
The spectral and SUSY properties for types (A1) -- (B2) are listed 
in the table in Appendix B.

Now we discuss how the previous works [\Boya, \GLR] on SUSY interval systems with the
Dirac $\delta(x)$-potentials fit in our general scheme.  In order to realize the boundary
condition that arises under the Dirac $\delta(x)$-potentials with different coupling
constants on the two lines, we consider the general ordered diagonal matrix $\bar D$ in
(\dbloblo) and
choose the conjugation matrix $V$ and the exchange matrix $S$ by
$V_1 = V_2 = e^{i{\pi
\over 4}\sigma_2}$ in (\vbloblo) and 
$S = Y$ given in (\esmtrx), respectively.  Then we find that the boundary condition (\bouzu)
becomes
$$
\psi^+_i(0) + 
{{L(\theta_i)}\over 2}\left( {\psi^+_i}^\prime(0) + {\psi^-_i}^\prime(0)\right) =0, 
\qquad
\psi^+_i(0)  = \psi^-_i(0), \qquad i = 1, \, 2.
\eqn\no
$$
These are indeed the conditions that we find under the potentials 
$V(x) =  g_i\delta(x)$ with strengths $g_i =
{{2\lambda^2}\over{L(\theta_i)}}$, and hence we see that  the pair of line
systems  having the Dirac
$\delta(x)$-potentials has an $N = 1$ SUSY for arbitrary strengths $g_1$ and $g_2$. 
In particular, if the two strengths are related by $g_1 = \pm g_2$, which occur when the
angles $\theta_i$ fulfill (\hondawake), then one can enhance the SUSY to $N = 4$ by
choosing $\mu$ as in (\hondawake).  The case discussed in [\Boya, \GLR, \Junker]
corresponds to
$g_1 = - g_2$, where the pair of systems is regarded as a Witten model with $N = 2$.  Our
analysis shows, however, that the number of SUSY can be doubled if one takes into the
exchange parity operations between the half lines, $x > 0$ and $x < 0$.

\ve

{\centerline {\bf 3. Two intervals with point singularity}}
\medskip

\secno=3 \meqno=1

In this section, we study systems consisting of two intervals, each given by
 $[-l,l]$ with a singular point at $x = 0$. 
As before, we split each of the intervals into two and thereby 
regard our Hilbert
space
${\cal H}$ as $L^2([0, l]) \otimes \C^4$ (see Fig.2).  Our Hamiltonian remains
to be the one in (\hamihami) shifted by $\mu^2$ so that (\ancc) holds.
Because of the walls at $x = l$ we now
have, we need to impose, in addition to the boundary
condition (\bouzu) at $x = 0$, an extra boundary condition at $x = l$. 
Let $U \in
U(4)$ be the characteristic matrix
for the condition at
$x = 0$ given in the block diagonal form (\bloblo), and 
similarly $U_l \in U(4)$ be the characteristic matrix
for the condition at
$x = l$.
Since the probability current must vanish at $x =
l$ separately on the branches
$\pm$, we require
$$
j^\pm_1(l) = 0 = j_2^\pm(l).
\eqn\concon
$$
Comparing with the previous case (\contcond), we realize that 
the characteristic matrix at the wall is diagonal, {\it i.e.}, 
$U_l = D_l$ where
$D_l$ is of the form (\monmon) with $\theta_i$ 
being replaced by the corresponding parameter $\theta_i^l \in [0, 2\pi)$.
With such $D_l$, the
boundary condition at $x = l$ is provided by
$$
(D_l-1)\Psi(l) + iL_0(D_l+I)\Psi^\prime(l) = 0. 
\eqn\lten
$$
Thus the two interval systems we are considering are characterized by
the matrix $U_{\rm tot} = U \times D_l \in U(2) \times U(1) \times U(1)$.

We now find systems that accommodate SUSY.  More explicitly, 
we seek a pair $(U_{\rm tot}, Q)$ with $U_{\rm tot} = U \times D_l$ and 
some supercharge $Q$
for which the  boundary conditions both at $x = 0$ and $x = l$ are compatible with the
SUSY transformations generated by
$Q$.  To this end, we first recall that, if the characteristic matrix 
$U$ is decomposed as (\udecomp), the supercharge $Q$ compatible with the 
boundary condition at $x = 0$
is given by (\gensc).  Thus, what remains to be seen is under what conditions this
supercharge
$Q$ is simultaneously compatible with the 
boundary condition at $x = l$.  As we did at the point $x = 0$ for the diagonal $D$, 
at the point $x =
l$ we also put 
$$
D_l = S^{-1}_l \bar{D}_l S_l,
\qquad
\bar D_l = \hbox{diag}(e^{i\theta_1^l}, e^{i\theta_2^l}, -1, -1),
\eqn\mokomoko
$$
with some appropriate exchange matrix $S_l$.
Then, by an analogous argument we see that the supercharge compatible 
with the boundary condition at $x=l$ is given by
$S^{-1}_lq(A_l, \mu_l;{K}_l)S_l$.
From the two compatibility conditions at $x = 0$ and
$x = l$, one has
$$
V^{-1}S^{-1}q(A, \mu;{K})SV
= Q = S^{-1}_lq(A_l, \mu_l;{K}_l)S_l,
\eqn\susysol
$$
which implies
$$
S_lV^{-1}S^{-1}\Gamma SVS^{-1}_l
=\Gamma_l,
\eqn\gamga
$$
and
$$
\mu S_lV^{-1}S^{-1}\Omega SVS^{-1}_l
=\mu_l\Omega_l.
\eqn\omom
$$
The total number of independent supercharges satisfying 
(\gamga) and (\omom) gives the number of the SUSY that the system possesses.
This will be seen by the number of the free parameters in $A$, and 
if we have
the full, four parameters, {\it i.e.}, if
$N = 4$, we may choose
$A = i\sigma_k$ and $A = {I_2}$ in (\susysol) 
to furnish a basis of supercharges $Q_i$, $i = 1, \ldots, 4$, as in
(\otohimesama) fulfilling the standard SUSY algebra (\ototachibana).

\topinsert
\epsfxsize 6.5cm
\ifx\omitpictures N  \centerline{\epsfbox {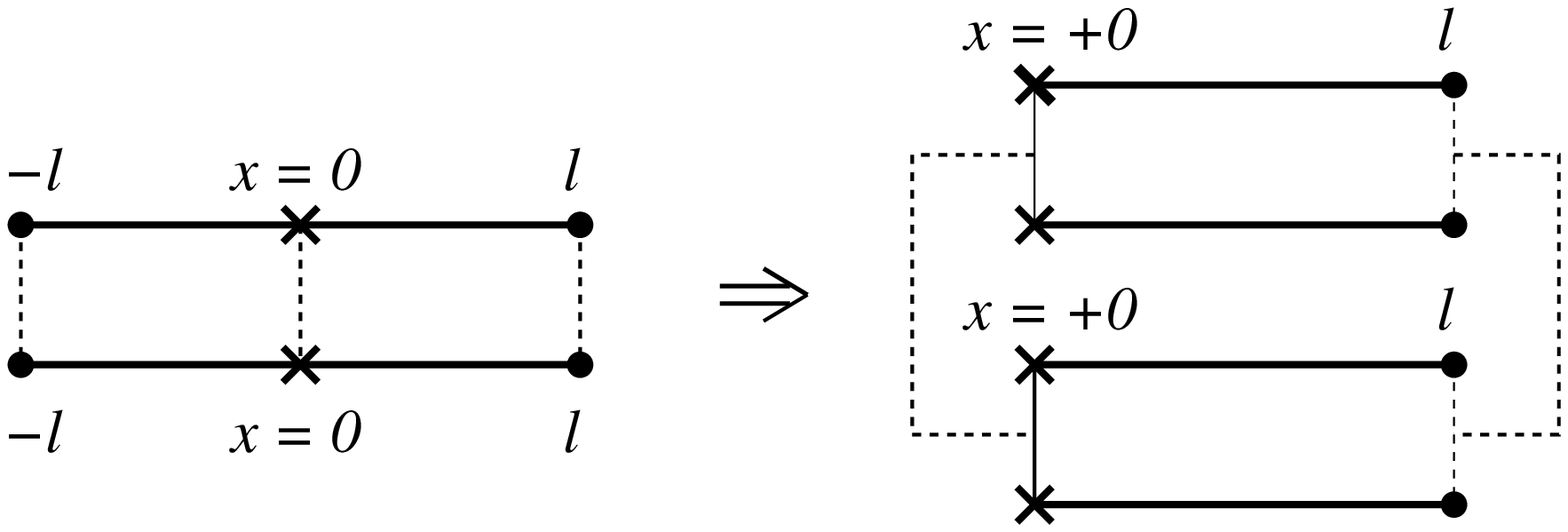}}  \fi
\abstract{{\bf Figure 2.}~A system of a pair of two intervals $[-l, l]$ each having a 
singularity at $x = 0$ may be identified with two systems of a pair of two half intervals
$(0, l]$ where the probability flow is allowed to pass between the two systems through $x =
0$.  The flow is not allowed at the other ends $x = l$.}
\bigskip
\endinsert

Because of the exchange matrices $S$ and $S_l$, our analysis becomes involved
compared to the previous case.  There are, however, two discrete operations 
which can be used to simplify our arguments.  These are the parity,
$$
{\cal P}: \quad
\Psi(x) \longrightarrow
({\cal P}\Psi)(x) := \Psi(l - x),
\eqn\no
$$
and the interchange,
$$
{\cal I}_X: \quad
\Psi(x) \longrightarrow
({\cal I}_X \Psi)(x) := X \Psi(x),
\eqn\no
$$
where $X$ is defined in (\esmtrx).  If systems that are connected by
these discrete operations are regarded to be essentially identical, 
there remain only the following 
four types of combinations for the diagonal matrices $(D, D_l)$:
$$
\eqalign{
\hbox{(a)}
\qquad
D &= \hbox{diag}(e^{i\theta_1}, e^{i\theta_2}, -1, -1), \quad D_l =
\hbox{diag}(e^{i\theta_1^l}, e^{i\theta_2^l}, -1, -1),
\cr
\hbox{(b)}
\qquad 
D &= \hbox{diag}(-1, -1, e^{i\theta_1}, e^{i\theta_2}), \quad D_l =
\hbox{diag}(e^{i\theta_1^l}, e^{i\theta_2^l}, -1, -1),
\cr
\hbox{(c)}
\qquad 
D &= \hbox{diag}(e^{i\theta_1}, -1, e^{i\theta_2}, -1), \quad D_l =
\hbox{diag}(e^{i\theta^l_1}, e^{i\theta^l_2}, -1, -1),
\cr
\hbox{(d)}
\qquad
D &= \hbox{diag}(e^{i\theta_1}, -1, e^{i\theta_2}, -1), \quad D_l =
\hbox{diag}(e^{i\theta_1^l}, -1, e^{i\theta_2^l}, -1).
}
$$

To proceed, except when we examine the implication of the general $B \ne 0$ case, 
we shall
restrict our analysis to the case $B = 0$, that is, we assume that at $x = 0$ the two
angle parameters $\theta_1$ and $\theta_2$ satisfy (\hondawake) and the constant $\mu$ is
chosen to be $\mu^2 = [\lambda/L(\theta)]^2$, and that similar conditions hold also at
$x = l$ 
(for which the angle $\theta^l_i$ will be used for $\theta_i$ in (\hondawake)).   This
restriction simplifies our argument considerably and ensures that all the allowed (at
most four) supercharges satisfy the standard orthogonal SUSY algebra (\ototachibana).  
Since (\omom) implies
$\mu_l^2=\mu^2$, we then have 
$\vert L(\theta_1) \vert =\vert L(\theta_2) \vert =\vert L(\theta_1^l)\vert =
\vert L(\theta_2^l)\vert$.  We thus have only one free scale parameter 
in all the boundary
conditions, and we shall specify it by the angle $\theta_1 = \theta$.  
With respect to the scale $L(\theta)$ set by $\theta$, we introduce the sign
functions to the remaining three scale parameters,
$$
s_2 = {{L(\theta_2)}\over{L(\theta)}}, 
\qquad 
s_1^l = {{L(\theta_1^l)}\over{L(\theta)}}, 
\qquad 
s_2^l = {{L(\theta_2^l)}\over{L(\theta)}}, 
\eqn\no
$$
which take either $1$ or $-1$.  
Among the possible choices for the signs of $\mu = \pm
\lambda/L(\theta)$ and 
$\mu_l= \pm
\lambda/L(\theta^l)$ (the choice does not affect
the supercharge
$Q$), for definiteness we choose 
$\mu = \mu_l = + \lambda/L(\theta)$.
With this choice the ${K}$ matrix at $x = 0$ and the corresponding 
${K}_l$ at $x = l$
defined similarly as (\amaterasu) become
$$
{K}= 
	\left( \matrix{
		1 & 0 \cr
		0 & s_2
	} \right), 
\qquad
{K}_l=
	\left( \matrix{
		s_1^l & 0 \cr
		0 & s_2^l
	} \right),
\eqn\lamdef
$$
that is, they are proportional to either ${I_2}$ or $\sigma_3$.
In order to solve (\gamga) and (\omom), 
we parametrize $A$ and $V_i$, $i=1,\, 2$ as
$$
A=e^{i\xi}e^{i\frac{\alpha}{2}
\sigma_3}e^{i\beta\sigma_2}e^{i\frac{\omega}{2}\sigma_3}, 
\qquad
V_i=e^{i\delta_i\sigma_2}e^{i\frac{\tau_i}{2}\sigma_3},
\eqn\vpara
$$
with $\xi,\alpha, \omega, \tau_i\in[0, 2\pi)$ and $\beta, \delta_i\in[0, \pi)$, and
consider the four types of the combinations, separately.

\bigskip
\noindent
{\bf 3.1. Type (a)} $\,\,\, D = \hbox{diag}(e^{i\theta_1}, e^{i\theta_2},
-1, -1), \quad
D_l =\hbox{diag}(e^{i\theta_1^l}, e^{i\theta_2^l}, -1, -1)$
\medskip

In this case, the rotation matrix $V_2$ 
becomes irrelevant for determining the characteristic matrix $U$, and 
we may choose 
$V_2 = {I_2}$ without loss of generality.
We then obtain three distinct solutions for (\gamga) and (\omom)
(for the detail, see Appendix A.1),
$$
\eqalign{
&\hbox{(a1)}
\qquad
s_2=s_1^l=s_2^l=1, \qquad\quad\,\,\quad\,\,\quad\,\,\quad\,\,\,\,\,\,\,\,\,\,\,
A_l=V_1^\dagger A,
\cr
&\hbox{(a2)}
\qquad
s_1^l=1, \quad s_2^l=s_2=-1, \quad
\delta_1=0, \,\,\,\,\,\,\quad A_l=V_1^\dagger A,
\cr
&\hbox{(a3)}
\qquad
s_1^l=s_2=-1, \quad s_2^l=1, \quad
\delta_1=\pi/2, \quad A_l=V_1^\dagger A.
}
$$
All the above three 
solutions admit the four supercharges 
$Q_i$ in (\otohimesama) and 
hence the systems possess an $N = 4$ SUSY.

In order to see what our SUSY systems are in more detail, let us first consider 
the type
(a1) solution which implies the boundary condition,
$$
\eqalign{
	&\psi_1^\pm(0) + L(\theta){\psi_1^\pm}^\prime(0)  = 0, \qquad
	\psi_1^\pm(l) + L(\theta){\psi_1^\pm}^\prime(l) = 0, \cr
	&\psi_2^\pm(0)  = 0, \qquad\qquad\qquad \qquad \psi_2^\pm(l) = 0.
}
\eqn\bdao
$$
{}For energy eigenstates of the Hamiltonian $H$ fulfilling (\bdao), we find 
the series
of eigenstates,
$$
\Phi_n(x) = 
	\left( \matrix{
		N_1(\sin k_nx - L(\theta)k_n\cos k_nx) \cr
		N_2(\sin k_nx - L(\theta)k_n\cos k_nx) \cr
		N_3\sin k_nx \cr
		N_4\sin k_nx
	} \right),
\eqn\no
$$
where $N_i$, $i = 1, \ldots, 4$, are arbitrary constants subject to the normalization
condition
$1 = \| \Phi_n \|^2 = \int^\infty_0 dx \vert \Phi_n(x)\vert^2$, and we have introduced
$$
k_n = {{n\pi}\over l}, \qquad n = 1, 2,\ldots.
\eqn\kzero
$$ 
Consequently, each of the energy levels is four-fold degenerate. 
In addition, we obtain the doubly degenerate ground states,
$$
\Phi_{\rm grd}(x) = 
	\left( \matrix{
		\tilde N_1 e^{-x/L(\theta)} \cr
		\tilde N_2 e^{-x/L(\theta)} \cr
		0 \cr
		0
	} \right),
\eqn\no
$$
with vanishing energy $E_{\rm grd} = 0$. 
These ground states are annihilated by any of the supercharges $Q_i$, $i = 1, \ldots,
4$, and  hence we see that the $N = 4$ SUSY of type (a1) is good (unbroken).

For type (a2), the boundary condition becomes
$$
\eqalign{
	&\psi_1^\pm(0) \pm L(\theta){\psi_1^\pm}^\prime(0)  = 0, \qquad
	\psi_1^\pm(l) \pm L(\theta){\psi_1^\pm}^\prime(l) = 0, \cr
	&\psi_2^\pm(0) = 0, \qquad \qquad \qquad \qquad \psi_2^\pm(l) = 0,
}
\eqn\no
$$
which admits the energy eigenstates (with $k_n$ given in (\kzero))
$$
\Phi_n(x) = 
	\left( \matrix{
		N_1(\sin k_nx - L(\theta)k_n\cos k_nx) \cr
		N_2(\sin k_nx + L(\theta)k_n\cos k_nx) \cr
		N_3\sin k_nx \cr
		N_4\sin k_nx
	} \right), 
\eqn\no
$$
which are four-fold degenerate.  As before, the ground states are 
$$
\Phi_{\rm grd}(x) = 
	\left( \matrix{
		\tilde N_1 e^{-x/L(\theta)} \cr
		\tilde N_2 e^{x/L(\theta)} \cr
		0 \cr
		0
	} \right),
\eqn\no
$$
with energy $E_{\rm grd} = 0$.  Again, 
these doubly degenerate ground states are annihilated by the supercharges 
$Q_i$,
and  hence type (a2) provides an $N = 4$ good SUSY, too.  
Type (a3) furnishes essentially the same 
$N = 4$ good SUSY system as type (a2), except that the upper two components of 
all the eigenstates 
$\Phi(x)$ are interchanged.

\bigskip
\noindent
{\bf 3.2. Type (b)} $\,\,\, 
D = \hbox{diag}(-1, -1, e^{i\theta_1}, e^{i\theta_2}),
\quad D_l =\hbox{diag}(e^{i\theta_1^l}, e^{i\theta_2^l}, -1, -1)$
\medskip

This time the rotation matrix $V_1$  
is irrelevant for specifying $U$ and hence we take  
$V_1 = {I_2}$. 
As for type (a), we obtain from (\gamga) and (\omom) the following
three solutions (see Appendix A.2):
$$
\eqalign{
&\hbox{(b1)}
\qquad
s_2=s_1^l=-1, \quad s_2^l=1, \quad
\quad\beta=0, \quad\quad\, A_l = A^\dagger V_2,
\cr
&\hbox{(b2)}
\qquad
s_2=s_2^l=-1, \quad s_1^l=1, \quad
\quad\beta=\pi/2, \quad A_l=A^\dagger V_2,
\cr
&\hbox{(b3)}
\qquad
s_2=1, \quad\quad\qquad s_1^l=s_2^l=-1, \qquad\quad\quad
A_l=A^\dagger V_2.
}
$$
This time, since $\beta$ is specified in Type (b1) and (b2), these two types admit only
two independent supercharges obtained, for instance,  by $A = i\sigma_3$ and $A = {I_2}$, and hence they possess an $N = 2$ SUSY.  Type (b3), on the other hand, admit
all the four supercharges $Q_i$ and hence has an
$N = 4$ SUSY.

{}For type (b1), the boundary condition becomes
$$
\eqalign{
&\psi_1^\pm(0) = 0, 
\qquad\quad\quad\quad
\psi_1^\pm(l) \mp L(\theta){\psi_1^\pm}^\prime(l) = 0, 
\quad\quad\quad\quad\quad \psi_2^\pm(l) = 0,\cr
&e^{\pm i\tau_2} \psi_2^\pm(0)  + \tan\delta_2\, {\psi_2^\mp}(0)  
+ L(\theta)\left( e^{\pm i\tau_2} {\psi_2^\pm}^\prime(0) \pm
\tan\delta_2\, {\psi_2^\mp}^\prime(0)\right) 
= 0.
}
\eqn\no
$$
We then find two distinct series of eigenstates; one given by 
$$
\Phi_n^{(1)}(x) = 
	\left( \matrix{
		N_1\sin k^{-}_nx \cr
		0 \cr
		N_2\cos\delta_2\sin k^{-}_n(x-l) \cr
		N_2\sin\delta_2 e^{i\tau_2}\sin k^{-}_n(x-l)
	} \right),
\eqn\no
$$
and the other by
$$
\Phi_n^{(2)}(x) = 
	\left( \matrix{
		0 \cr
		N_3\sin k^{+}_nx \cr
		-N_4\sin\delta_2\sin k^{+}_n(x-l) \cr
		N_4\cos\delta_2e^{i\tau_2}\sin k^{+}_n(x-l)
	} \right),
\eqn\no
$$
where discrete $k_n^{\pm} = k_n^{\pm}(\theta) > 0$, $n = 1, \, 2, \, 3\ldots$, 
are obtained as the solutions of 
$$
L(\theta)k_n^{\pm} \pm \tan (k_n^{\pm}l)= 0.
\eqn\kpm
$$
{}For $0<L(\theta)<l$ we have the 
additional eigenstates, 
$$
\Phi^{(1)}_{\rm grd}(x) = 
	\left( \matrix{
		\tilde N_1\sinh \kappa^{-}x \cr
		0 \cr
		\tilde N_2 \cos\delta_2\sinh \kappa^{-}(x-l) \cr
		\tilde N_2 \sin\delta_2 e^{i\tau_2}\sinh \kappa^{-}(x-l)
	} \right),
\eqn\fstgrd
$$
while for
$-l<L(\theta)<0$ we obtain
$$
\Phi^{(2)}_{\rm grd}(x) = 
	\left( \matrix{
		0 \cr
		\tilde N_3\sinh \kappa^{+}x \cr
		-\tilde N_4\sin\delta_2 \sinh \kappa^{+}(x-l) \cr		
		\tilde N_4\cos\delta_2e^{i\tau_2}\sinh \kappa^{+}(x-l)
	} \right),
\eqn\secgrd
$$
where 
$\kappa^{\pm}=\kappa^{\pm}(\theta) > 0$ are the solutions of 
$$
L(\theta)\kappa_n^{\pm} \pm \tanh (\kappa_n^{\pm}l)= 0.
\eqn\kppm
$$
These provide the ground state of the system with energy
$$
E_{\rm grd}^\pm = -{{\hbar^2(\kappa^{\pm})^2}\over{2m}}
+ \left({\lambda}\over{L(\theta)}\right)^2.
\eqn\izumo
$$
{}For $L(\theta) = l$, the state $\Phi^{(1)}_{\rm grd}$ 
reduces to 
$$
\Phi^{(1)}_{\rm grd}(x) =
	\left( \matrix{
		\bar N_1 \enspace x \cr
		0 \cr
		\bar N_2\cos\delta_2 \, (x-l) \cr
		\bar N_2\sin\delta_2 \,e^{i\tau_2} \, (x-l)
	} \right),
\eqn\zst
$$
while for $L(\theta) = -l$, $\Phi^{(2)}_{\rm grd}$ reduces to
$$
\Phi^{(2)}_{\rm grd}(x) =
	\left( \matrix{
		0 \cr
		\bar N_3 \enspace x \cr
		-\bar N_4\sin\delta_2  \, (x-l) \cr
		\bar N_4\cos\delta_2 \, e^{i\tau_2} \, (x-l)
	} \right).
\eqn\zstt
$$
These eigenstates are all doubly degenerate irrespective of the energy, and are
related by the SUSY transformations generated by the two supercharges mentioned
above.  We also note that the ground state energy is positive
$E_{\rm grd}^\pm > 0$ and hence the $N = 2$ SUSY of the system
is broken.  Type (b2) is essentially the same as (b1), and we will omit to give its
detail here.

For type (b3), the boundary condition is 
$$
\eqalign{
&\psi_1^\pm(0) = 0, \qquad  
\psi_1^\pm(l) - L(\theta){\psi_1^\pm}^\prime(l) =0, \cr
&\psi_2^\pm(0) + L(\theta){\psi_2^\pm}^\prime(0) = 0, 
\qquad \psi_2^\pm(l) =0. \cr
}
\eqn\no
$$
The energy eigenstates are 
$$
\Phi_n(x) = 
	\left( \matrix{
		N_1 \sin k_n^- x \cr
		N_2\sin k_n^- x \cr
		N_3\sin k_n^- (x-l) \cr
		N_4\sin k_n^- (x-l)
	} \right),
\eqn\pest
$$
where $k_n^-  > 0$ are given in (\kpm). 
Besides, for $0<L(\theta)<l$, we have the ground 
states provided by (\pest) with the replacement $k_n^- \rightarrow i\kappa^-$,
that is, $\kappa^- > 0$ is the solution of (\kppm).  
{}Further, if $L(\theta) = l$, 
the states (\pest) with the sine functions 
formally 
replaced by their arguments $\sin z \rightarrow z$ become eigenstates.
All of these eigenstates are four-fold degenerate, and
they are related by the SUSY transformations generated by $Q_i$. 
As before, the $N =
4$ SUSY is broken in the system.

At this point, we briefly mention the $B \ne 0$ case for
which the supercharges (\izanagi) do not necessarily satisfy the standard
orthogonal SUSY algebra (\ototachibana).  
For simplicity, we only consider the special case where we have
$V_1 = V_2 = {I_2}$ and choose $B$ by
(\yamatotakeru). 
Here we can readily solve (\gamga) and
(\omom) to obtain, for instance, the solution,
$$
\eqalign{
&L(\theta_1^l)=-L(\theta_1), 
\quad L(\theta_2^l)=-L(\theta_2), \quad \beta=\beta_l=0, \cr
&\alpha_l+\omega_l=\alpha+\omega, 
\quad \xi_l=\xi, \quad \mu_l=\pm\mu, \quad b_1^l=\mp b_1, \quad b_2^l=\mp b_2.
}
\eqn\gespsol
$$
The corresponding supercharges may then be provided by 
$q({I_2}, \mu; {K})$ 
and $q(i\sigma_3, \mu;{K})$, and the boundary condition
becomes
$$
\eqalign{
&\psi_1^\pm(0) = 0, \qquad \psi_1^\pm(l) - L(\theta_1){\psi_1^\pm}^\prime(l) =0,
\cr 
&\psi_2^\pm(0)  + L(\theta_1){\psi_2^\pm}^\prime(0) = 0, 
\qquad \psi_2^\pm(l) =0. \cr
}
\eqn\no
$$
Because of the two scale parameters $L(\theta_1)$ and $L(\theta_2)$ we have
now, the energy eigenstates consist of the two series,
$$
\Phi^{(1)}_n(x) = 
	\left( \matrix{
		N_1 \sin k^{-}_n(\theta_1) x \cr
		0 \cr
		N_3\sin k^{-}_n(\theta_1)(x-l) \cr
		0
	} \right), \qquad
\Phi^{(2)}_n(x)=
	\left( \matrix{
		0 \cr
		N_2\sin k^{-}_n(\theta_2)x \cr
		0 \cr
		N_4\sin k^{-}_n(\theta_2)(x-l)
	} \right),
\eqn\bexcs
$$
where the discrete $k^{-}_n(\theta_i) > 0$, $i=1,2$, are given by (\kpm)
with $\theta = \theta_i$. 
Similarly, for $0<L(\theta_i)<l$ there arise ground
states obtained by the replacement $k_n^{-} \rightarrow i\kappa^{-}$ in (\bexcs)
with energy $E_{\rm grd} = -\hbar^2[\kappa^{-}(\theta_i)]^2/(2m) + [\lambda/L(\theta_i)]^2 >
0$.  As the scale parameters approach
$L(\theta_i) = l$ these solutions reduce to states analogous to (\zst) or (\zstt).  
All of these states are doubly degenerate and are
related by the SUSY transformations generated by the two supercharges,
implying that the $N = 2$ SUSY of the system is broken.

In the examples discussed above, we observe that the spectrum of the system consists of
one or two \lq regular\rq{} series of levels specified by $k_n$ or $k_n^\pm$ 
with integers $n$,
plus a few
\lq isolated\rq{} levels some of which may become the ground states.  Obviously, these
correspond to states with positive and negative energies, respectively, if the constant energy
shift in the Hamiltonian is absent.   The same spectral composition will be seen in
all the examples we shall discuss below.

\bigskip
\noindent
{\bf 3.3. Type (c)} $\,\,\, 
D = \hbox{diag}(e^{i\theta_1}, -1, e^{i\theta_2}, -1),
\quad D_l =\hbox{diag}(e^{i\theta^l_1},
e^{i\theta^l_2}, -1, -1)$
\medskip

The conditions (\gamga) and (\omom) are met again by three distinct types of
solutions (see Appendix A.3):
$$ 
\eqalign{
\hskip-.1cm
\hbox{(c1)}
\qquad 
&s_2=s_1^l=1, \quad s_2^l=-1, 
\quad \delta_1=0, \quad \beta=\pi/2, \quad
\beta_l=\pi/2-\delta_2,\quad \omega_l=\tau_2\pm\pi,
\cr
&\xi_l=(\alpha-\omega)/2, \quad 
\alpha_l=2\xi-\tau_1\pm\pi \quad 
\hbox{or}
\quad \xi_l=(\alpha-\omega)/2+\pi, \quad
\alpha_l=2\xi-\tau_1\mp\pi,\cr
\hbox{(c2)}
\qquad
&s_2=s_2^l=1, \quad s_1^l=-1,
\quad \delta_1= \pi/2, \quad \beta=\pi/2, \quad
\beta_l=\delta_2,\quad \omega_l=\tau_2, \cr
&\xi_l=(\alpha-\omega)/2, \quad \alpha_l=-2\xi-\tau_1 \quad 
\hbox{or} 
\quad 
\xi_l=(\alpha-\omega)/2+\pi, \quad
\alpha_l=2\pi-2\xi-\tau_1, \cr
\hbox{(c3)}
\qquad
&s_2=-1, \quad s_1^l=s_2^l=1, \quad
\beta=\pi/2,
}
$$
where in (c3) the remaining parameters $(\xi_l, \alpha_l, \beta_l,
\omega_l)$ and
$(\xi,
\alpha, \omega, \delta_i, \tau_i)$ are determined from (\gamga). 
All of these types have an $N = 2$ SUSY, because they possess two independent
supercharges.  Explicitly, they may be chosen to be the pairs, 
$q(e^{i(\delta_2-\frac{\pi}{2})\sigma_2} e^{i\frac{\tau_2}{2}\sigma_3}, \mu;{K})$ 
and
$q(i\sigma_3e^{i(\delta_2-\frac{\pi}{2})\sigma_2}
e^{i\frac{\tau_2}{2}\sigma_3}, \mu;{K})$ for (c1), 
$q(e^{i\delta_2\sigma_2}e^{i\frac{\tau_2}{2}\sigma_3}, \mu; {K})$ and
$q(i\sigma_3e^{i\delta_2\sigma_2}e^{i\frac{\tau_2}{2}\sigma_3}, \mu; {K})$ for (c2),
and
$V^{-1}Y^{-1}q(i\sigma_1, \mu;{K})YV$ and $V^{-1}Y^{-1}q(i\sigma_2,
\mu;{K})YV$ for (c3), respectively.

{}For type (c1), the boundary condition reads
$$
\eqalign{
&\psi_1^+(0) + L(\theta){\psi_1^+}^\prime(0) = 0, 
\quad \psi_1^-(0) = 0, \quad \psi_1^\pm(l)  \pm L(\theta){\psi_1^\pm}^\prime(l) =
0, 
\cr 
&e^{i\tau_2}\psi_2^+(0) +
\tan\delta_2\psi_2^-(0) + 
L(\theta)\left(e^{i\tau_2}{\psi_2^+}^\prime(0) +
\tan\delta_2{\psi_2^-}^\prime(0)\right) = 0, \cr
&e^{-i\tau_2}\psi_2^-(0)-\tan\delta_2\psi_2^+(0) = 0, 
\qquad\qquad\qquad\qquad\qquad\,\, \psi_2^\pm(l) = 0, }
\eqn\no
$$
which admits two regular series of eigenstates given by
$$
\Phi^{(1)}_n(x) = 
	\left( \matrix{
		N_1\bigl(\sin k_n x -
		L(\theta)k_n\cos k_n x \bigr) \cr
		0 \cr
		-N_2\sin\delta_2\sin k_n(x-l) \cr
		N_2\cos\delta_2e^{i\tau_2}\sin k_n(x-l)
	} \right), 
\eqn\eiei
$$
and 
$$
\Phi^{(2)}_n(x) = 
	\left( \matrix{
		0 \cr
		N_3\sin k^{-}_nx \cr
		N_4\cos\delta_2 \sin k^{-}_n(x-l) \cr
		N_4\sin\delta_2 e^{i\tau_2} \sin k^{-}_n(x-l)
	} \right).
\eqn\eit
$$
{}For $0<L(\theta)<l$, we have the
isolated eigenstate
$$
\Phi^{(1)}(x) = \Phi_{\rm grd}(x) =
	\left( \matrix{
		\tilde N_1e^{-x/L(\theta)} \cr
		0 \cr
		0 \cr
		0
	} \right),
\eqn\eig
$$
which is the ground state with $E_{\rm grd} = 0$ and
$$
\Phi^{(2)}(x) = 
	\left( \matrix{
		0 \cr
		\tilde N_3\sinh \kappa^{(2)}x \cr
		\tilde N_4\cos\delta_2 \sinh \kappa^{(2)}(x-l) \cr
		\tilde N_4\sin\delta_2 e^{i\tau_2} \sinh \kappa^{(2)}(x-l)
	} \right),
\eqn\eitg
$$
with $E = -\hbar^2({\kappa^{-}})^2/(2m) + [\lambda/L(\theta_i)]^2 > 0$. 
{}For $L(\theta)=l$, this reduces to
$$
\Phi^{(2)}(x) = 
	\left( \matrix{
		0 \cr
		\tilde N_3 \enspace x \cr
		\tilde N_4\cos\delta_2 \enspace (x-l) \cr
		\tilde N_4\sin\delta_2 e^{i\tau_2}\enspace (x-l)
	} \right).
\eqn\eitz
$$
These eigenstates may be classified into the two series; one is a \lq good SUSY series\rq{}
given by (\eiei) and (\eig) and the other is a \lq broken SUSY series\rq{} given by (\eit),
(\eitg) and (\eitz).

Type (c2) provides a system which is analogous to (c1), and we shall not present the
content of the system.  For type (c3), on the other hand, the boundary
condition is given by
$$
\eqalign{
&e^{i\tau_1}\psi_1^+(0) + \tan\delta_1\psi_1^-(0)
+ L(\theta)\left(e^{i\tau_1}{\psi_1^+}^\prime(0) + \tan\delta_1{\psi_1^-}^\prime(0)\right)
= 0, \cr 
&e^{-i\tau_1}\psi_1^-(0)-\tan\delta_1\psi_1^+(0) = 0, \,\,\qquad\qquad  
\psi_1^\pm(l)  + L(\theta){\psi_1^\pm}^\prime(l) = 0, \cr
&e^{i\tau_2}\psi_2^+(0) +
\tan\delta_2\psi_2^-(0)
- L(\theta)\left(e^{i\tau_2}{\psi_2^+}^\prime(0) +
\tan\delta_2{\psi_2^-}^\prime(0)\right) 
= 0, \cr 
&e^{-i\tau_2}\psi_2^-(0)-\tan\delta_2\psi_2^+(0) = 0, \qquad\qquad \qquad \qquad \qquad 
\,\, 
\psi_2^\pm(l) = 0. }
\eqn\no
$$
The eigenstates are then
$$
\Phi^{(1)}_n(x) = 
	\left( \matrix{
		N_1 \cos\delta_1
		\bigl(\sin k_nx - L(\theta)k_n\cos k_nx\bigr) \cr
		N_1 \sin\delta_1e^{i\tau_1}
		\bigl(\sin k_nx - L(\theta)k_n\cos k_nx\bigr) \cr
		-N_2\sin\delta_2\sin k_n(x-l) \cr
		N_2\cos\delta_2e^{i\tau_2}\sin k_n(x-l)
	} \right),
\eqn\aei
$$
and
$$
\Phi^{(2)}_n(x) = 
	\left( \matrix{
		-N_3 \sin\delta_1\sin k^{+}_nx \cr
		N_3 \cos\delta_1e^{i\tau_1}\sin k^{+}_nx \cr
		N_4 \cos\delta_2 \sin k^{+}_n(x-l) \cr
		N_4 \sin\delta_2 e^{i\tau_2} \sin k^{+}_n(x-l)
	} \right).
\eqn\bei
$$
We have the isolated eigenstate
$$
\Phi^{(1)}(x) = \Phi_{\rm grd}(x) =
	\left( \matrix{
		\tilde N_1 \cos\delta_1 e^{-x/L(\theta)} \cr
		\tilde N_1 \sin\delta_1e^{i\tau_1}e^{-x/L(\theta)} \cr
		0 \cr
		0
	} \right),
\eqn\aeig
$$
which is the ground state with $E_{\rm grd} = 0$.  Also, for
$-l<L(\theta)<0$, we have additionally 
$$
\Phi^{(2)}(x) = 
	\left( \matrix{
		-\tilde N_3 \sin\delta_1\sinh \kappa^{+}x \cr
		\tilde N_3 \cos\delta_1e^{i\tau_1}\sinh \kappa^{+}x \cr
		\tilde N_4 \cos\delta_2 \sinh \kappa^{+}(x-l) \cr
		\tilde N_4 \sin\delta_2 e^{i\tau_2} \sinh \kappa^{+}(x-l)
	} \right),
\eqn\beig
$$
with $E = -\hbar^2(\kappa^{+})^2/(2m) + [\lambda / L(\theta)]^2 > 0$. 
{}For $L(\theta)=-l$, this reduces to a state obtained similarly as (\eitz) 
from (\eitg).
Again, these eigenstates are classified into two 
series; one is a good SUSY series given by (\aei) and (\aeig) and the other is a broken
SUSY series given by (\bei) and (\beig).  For 
the special choice $\delta_1=\delta_2=\pi/4$ and
$\tau_1=\tau_2=0$, the resultant system (formulated in the original Hilbert space ${\cal
H}=L^2([-l,l])\otimes\C^2$) turns out to be the SUSY system obtained under 
the attractive and repulsive pair of the Dirac
$\delta(x)$-potentials
$V(x) = \mp\frac{2\lambda^2}{L(\theta)}\delta(x)$.

\bigskip
\noindent
{\bf 3.4. Type (d)} $\,\,\, 
D = \hbox{diag}(e^{i\theta_1}, -1, e^{i\theta_2}, -1),
\quad D_l =\hbox{diag}(e^{i\theta_1^l}, -1, e^{i\theta_2^l}, -1)$
\medskip

We here find from (\gamga) and (\omom) six distinct types of solutions (see Appendix
A.4).  Among them, four are
$$
\eqalign{
\hskip-.1cm
\hbox{(d1)}
\qquad 
&s_1^l=1, \quad s_2^l=-s_2, \quad \delta_1=0, 
\quad \delta_2=\pi/2, \quad \beta=\beta_l=0, \cr
&\xi_l=(\alpha+\omega-\tau_1+\tau_2\pm\pi/2)/2, 
\quad \alpha_l+\omega_l=2\xi-\tau_1+\tau_2\mp\pi/2,\cr
\hbox{(d2)}
\qquad 
&s_1^l=-1, \quad s_2^l=s_2, \quad \delta_1=\pi/2, 
\quad \delta_2=0, \quad \beta=\beta_l=0, \cr
&\xi_l=-(\alpha+\omega+\tau_1+\tau_2\mp\pi/2)/2, 
\quad \alpha_l+\omega_l=-2\xi-\tau_1+\tau_2\pm\pi/2,\cr
\hbox{(d3)}
\qquad 
&s_1^l=1, \quad s_2=s_2^l=-1, \quad \delta_1 = \delta_2 \neq 0 \enspace{\rm or}\enspace
\pi/2, 
\quad \beta = \beta_l = \pi/2, \cr
&\xi=0\enspace{\rm or}\enspace\pi, \quad \xi_l=-(\tau_1+\tau_2)/2+\xi, 
\quad \alpha_l-\omega_l=\alpha-\omega,
\cr
\hbox{(d4)}
\qquad 
&s_1^l=1, \quad s_2=s_2^l=-1, \quad \delta_2 = \pi-\delta_1 \neq \pi/2, 
\quad \beta = \beta_l = \pi/2, \cr
&\xi=\pi/2 \enspace {\rm or}\enspace 3\pi/2, 
\quad \xi_l=-(\tau_1+\tau_2)/2, \quad
\alpha_l-\omega_l=\alpha-\omega. 
}
$$
All of these have an $N = 2$ SUSY, where the supercharges may be given by 
$Y^{-1}q({I_2};{K})Y$ and $Y^{-1}q(i\sigma_3;{K})Y$ for type (d1) and (d2), 
or by $Y^{-1}q(ie^{-i(\tau_1+\tau_2)/2}\sigma_1;{K})Y$ and \hfil\break 
$Y^{-1}q(ie^{-i(\tau_1+\tau_2)/2}\sigma_2;{K})Y$ for type (d3) and (d4).
The remaining two are
$$
\eqalign{
\hskip-.1cm
\hbox{(d5)}
\qquad 
&s_1^l=1, \quad s_2^l=s_2, \quad \delta_1 = \delta_2 = 0, \quad \beta_l = \beta, 
\quad \xi_l=\xi-(\tau_1+\tau_2)/2, \cr
&\alpha_l=\alpha-(\tau_1-\tau_2)/2, 
\quad \omega_l=\omega-(\tau_1-\tau_2)/2, 
\cr
\hbox{(d6)}
\qquad 
&s_2=1, \quad s_1^l=s_2^l=-1, 
\quad \delta_1 = \delta_2 = \pi/2, \quad \beta_l = \beta, \quad
\xi_l=-\xi-(\tau_1+\tau_2)/2, 
\cr
&\alpha_l=-\omega-(\tau_1-\tau_2)/2\pm\pi/2, \quad
\omega_l=-\alpha-(\tau_1-\tau_2)/2\pm\pi/2, 
}
$$
which have four supercharges $Y^{-1}Q_iY$, $i = 1,2,3,4$, 
and hence possess an $N = 4$ SUSY. 

For types (d1) and (d2), one observes that under the SUSY transformations
the eigenstates exchange either the upper two or the lower two
components.  This implies  that these systems are essentially the sum of
two disconnected single lines with a point singularity, and hence reduce to the systems
considered earlier in [\TUIT].  Type (d3), on the other hand, provides a novel
SUSY system, and we here mention only the simple case
$\theta = 0$.  The boundary condition then becomes
$$
\eqalign{
e^{-i\tau_i}\psi_i^-(0)-\tan\delta_1\psi_i^+(0) &= 0, 
\qquad 
e^{i\tau_i}{\psi_i^+}^\prime(0)+\tan\delta_1{\psi_i^-}^\prime(0)=0, 
\cr
{\psi_i^+}^\prime(l)&=0, 
\qquad \psi_i^-(l)=0,
\qquad i = 1, \, 2,
}
\eqn\no
$$
and the eigenstates are 
$$
\eqalign{
&\Phi^{(1)}_n(x) =
	N_1\left( \matrix{
		-\cos \bar k_n(x-l) \cr
		e^{i\tau_1}\sin \bar k_n(x-l) \cr
		0 \cr
		0
	} \right),
\qquad
\Phi^{(2)}_n(x) =
	N_2\left( \matrix{
		0 \cr
		0 \cr
		-\cos \bar k_n(x-l) \cr
		e^{i\tau_2}\sin \bar k_n(x-l),
	} \right), 
}
\eqn\nnei
$$
where $\bar k_n= k_n +\delta_1/l$ for $n\in\Z$. 
Unless $\delta_1 = 0$ or $\pi/2$, all eigenstates are 
doubly degenerate and 
related under SUSY transformations generated by the two supercharges, implying that
the $N = 2$ SUSY is broken.  For $\delta_1 = 0$, except for the ground states which
are doubly degenerate, all excited states are four-fold degenerate, and
for $\delta_1 = \pi/2$, all eigenstates are four-fold degenerate (see
Fig.3).

\topinsert
\epsfxsize 6.5cm
\ifx\omitpictures N  \centerline{ \epsfbox {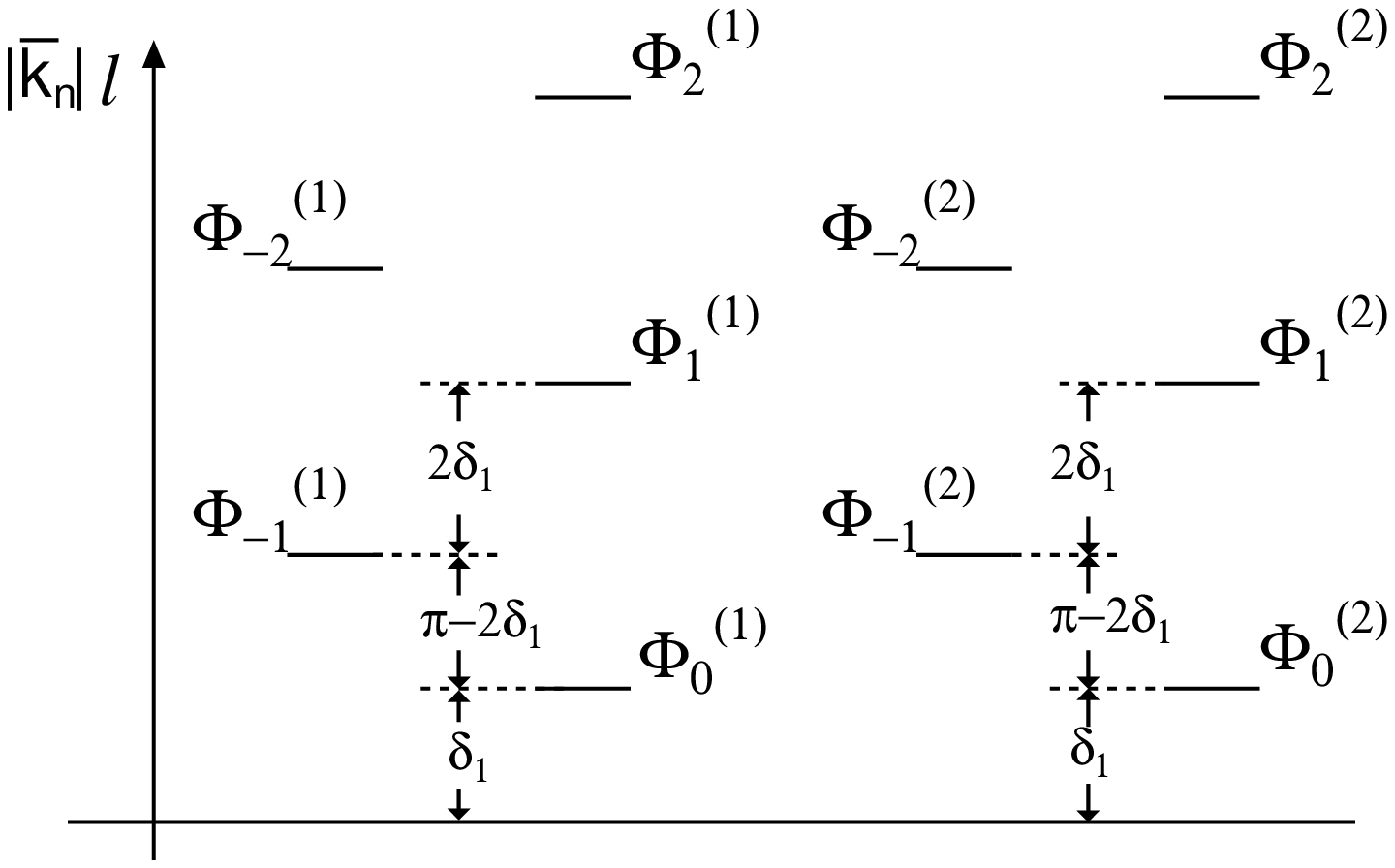} }  \fi
\abstract{{\bf Figure 3.}~Energy levels in the $N = 2$ SUSY system of type (d3) for $\theta =
0$.   All the levels are doubly degenerate unless $\delta_1 = 0$ or $\pi/2$.}
\bigskip
\endinsert

Type (d4) has the boundary condition 
$$
\eqalign{
\psi_1^+(0) + L(\theta){\psi_1^+}^\prime(0) &= 0, \qquad
\psi_1^+(l)+ L(\theta){\psi_1^+}^\prime(l) = 0, \quad
\psi_1^-(0) = \psi_1^-(l) = 0,
\cr
s_2\psi_2^+(0) + L(\theta){\psi_2^+}^\prime(0) &= 0, 
\quad 
s_2\psi_2^+(l) + L(\theta){\psi_2^+}^\prime(l) = 0, 
\quad \psi_2^-(0) =
\psi_2^-(l) = 0, 
}
\eqn\no
$$
and the regular series of eigenstates are
$$
\Phi_n(x) = 
	\left( \matrix{
		N_1(\sin k_nx - L(\theta)k_n\cos k_nx) \cr
		N_2 \sin k_nx \cr
		N_3 (\sin k_nx - s_2L(\theta)k_n\cos k_nx)\cr
		N_4\sin k_nx
	} \right), 
\eqn\no
$$
The ground states are then found to be
$$
\Phi_{\rm grd}(x) = 
	\left( \matrix{
		\tilde N_1 e^{-x/L(\theta)} \cr
		0 \cr
		\tilde N_2 e^{-s_2 x/L(\theta)} \cr
		0
	} \right),
\eqn\no
$$
with $E_{\rm grd} = 0$. 
The energy levels in the regular series are four-fold degenerate, while
the ground states are doubly degenerate.  The ground states are annihilated by the
four supercharges and, hence, the 
$N = 4$ SUSY is good.  When $\theta=0$, the system becomes equivalent to
one given by
$\delta_1=0$ in (\nnei).

{}For type (d6), the boundary condition reads
$$
\psi_i^+(0) = 0, 
\quad 
\psi_i^-(0) + L(\theta){\psi_i^-}^\prime(0) = 0,
\quad 
\psi_i^+(l) - L(\theta){\psi_i^+}^\prime(l) = 0, \quad \psi_i^-(l) =0, 
\eqn\no
$$
for $i = 1$, 2, and 
the regular eigenstates are 
$$
\Phi_n(x) = 
	\left( \matrix{
		N_1 \sin k^-_nx \cr
		N_2\sin k^-_n(x-l) \cr
		N_3\sin k^-_nx \cr
		N_4\sin k^-_n(x-l)
	} \right).
\eqn\typeds
$$
As before, for $0<L(\theta)<l$, isolated degenerate 
eigenstates are
obtained by 
$k^-_n \rightarrow i\kappa^-$ in (\typeds)
with $E > 0$.
All eigenstates, including the ground states, are four-fold degenerate and related by the SUSY
transformations, and hence the $N = 4$ SUSY is broken.  For $\theta=0$, the system coincides
with one given by
$\delta_1=\pi/2$ in (\nnei). 
These types of systems (a1) -- (d6) discussed above are listed in Appendix B.

{}Finally, we mention that a yet further extension of SUSY systems from those
considered in the present paper may be realized for restricted systems by incorporating
the possibility of  separate flips
of components of states.   Namely, for a state 
$\Psi(x) = (\psi_1^+(x), \psi_1^-(x), \psi_2^+(x), \psi_2^-(x))^T$ 
we may consider a number of discrete
transformations which flip each of the four
components, separately.  One of them is defined by
$$
{\cal F}^+_1:\quad
\Psi(x) \longrightarrow
({\cal F}^+_1\Psi)(x)=
	\left( \matrix{
		\psi_1^+(l-x) \cr
		\psi_1^-(x) \cr
		\psi_2^+(x) \cr
		\psi_2^-(x) 
	} \right).
\eqn\no
$$
This and other similarly defined flip operators, 
${\cal F}^-_1$, ${\cal F}^+_2$ and ${\cal F}^-_2$, are well-defined 
for systems for which no probability current flow is allowed at the
singularities.  Such systems occur when the characteristic matrix $U$ is
diagonal $U = D$, namely, 
when $U_{\rm tot} = D \times D_l \in [U(1)]^4$ for which the
four operators, which fulfill
$[{\cal F}^\pm_i]^2=\hbox{id}_{\cal H}$, induce the
transformations 
$U_{\rm tot} \rightarrow U_{\rm tot}^{{\cal F}^\pm_i}$ where $U_{\rm tot}^{{\cal
F}^\pm_i}$ is given by the exchange of the corresponding diagonal components 
between $D$ and
$D_l$.  Now, if the pair $(U_{\rm tot}, Q)$ is a SUSY system, then clearly the
new pair $(U_{\rm tot}^{{\cal F}^\pm_i}, {\cal F}^\pm_i Q [{\cal F}^\pm_i]^{-1})$
provides also a SUSY system. For example, if we implement this extension to 
type (b3), we find three novel SUSY systems characterized by
$$
\eqalign{
D^{(1)} &= \hbox{diag}(e^{i\theta}, -1, e^{i\theta}, e^{i\theta}),
\qquad D_l^{(1)} =\hbox{diag}(-1, e^{-i\theta}, -1, -1), \cr
D^{(2)} &= \hbox{diag}(e^{i\theta}, e^{i\theta}, e^{i\theta}, e^{i\theta}),
\qquad D_l^{(2)} =\hbox{diag}(-1, -1, -1, -1), \cr
D^{(3)} &= \hbox{diag}(e^{i\theta}, e^{i\theta}, -1, e^{i\theta}),
\qquad D_l^{(3)} =\hbox{diag}(-1, -1, e^{-i\theta}, -1), \cr
}
\eqn\no
$$
where we have used ${\cal F}^+_1$ for the first, and the combinations
${\cal F}^+_1{\cal F}^-_1$ and ${\cal F}^+_1{\cal F}^-_1{\cal F}^+_2$
for the second and the third, respectively.
The spectral properties of these are, of
course, the same as the original systems.

\ve
\centerline{ {\bf 4. Self-adjointness of the supercharge } }
\medskip

\secno=4 \meqno=1

In this section we wish to address the question of the self-adjointness of the
supercharge 
$Q = q(A, \mu; {K})$ in (\izanagi) for systems of 
two intervals discussed in section 3.  If $Q$ is a self-adjoint operator, then 
for any state $\Psi(x)$ belonging to its
domain 
$\Psi \in {\cal D}(Q) \subset {\cal H} \simeq L^2((0, l])\otimes\C^4$ 
we have
$$
\int_{0}^l\Psi^\dagger(x)\, (Q\Psi)(x)\, dx = \int_{0}^l(Q\Psi)^\dagger(x)
\, \Psi(x) \,dx, 
\eqn\marimo
$$
and also ${\cal D}(Q^\dagger) = {\cal D}(Q)$ for the adjoint $Q^\dagger$ of $Q$.
Since the $\mu$-term in the supercharge (\izanagi) is regular, it drops out
from the above condition (\marimo) leaving only the first derivative term there.
We may thus consider the simpler supercharge $Q =
-i\lambda\frac{d}{dx}\otimes\Gamma$ in finding possible domains for ${\cal D}(Q)$
below, based on the theory of self-adjoint extension [\RS].

To start, let us consider an operator $Q_0$ given in the same differential
form as $Q$ but defined on the domain 
$$
{\cal
D}(Q_0)=\left\{
\Psi \,\big\vert\,
\Psi(x)
\in \hbox{AC}[(0,l])\otimes\C^4,\, \,(Q_0\Psi)(x) \in {\cal H}, \,
\,\Psi(0) = \Psi(l)= 0
\right\},
\eqn\no
$$
where $\hbox{AC}[(0,l])$ is the space of absolutely continuous functions 
on $(0,l]$.  Clearly, its adjoint operator $Q^\dagger_0$ has also the same
operator form as
$Q$ and has the domain
${\cal D}(Q^\dagger_0) = \{\eta \,| \,\eta(x)\in {\cal H}, (Q^\dagger_0\eta)(x)\in
{\cal H}\}$. 
Now we 
consider eigenvectors $v_m$ (with eigenvalues $a_m$) of $\Gamma$, {\it
i.e.}, 
$\Gamma v_m = a_m v_m$ for $m = 1 \ldots, 4$, and thereby decompose any state
$\Psi(x)$ as
$$
\Psi(x)= \sum_{m=1}^{4}g_m(x)\, \upsilon_m,
\eqn\pdecomp
$$
where $g_m(x)$ are coefficient functions.   To implement the programme of
self-adjoint extension for $Q_0$, one needs to find the solutions for
$Q^\dagger_0\Psi_{\pm i} = -i\lambda\frac{d}{dx}\otimes\Gamma\Psi_{\pm i}(x) = \pm
i\Psi_{\pm i}(x)$, which, under the decomposition (\pdecomp), becomes
$$
-i\lambda\sum_{m=1}^4a_m \frac{d}{dx}g_m(x)\,\upsilon_m
=\pm i\sum_{m=1}^4g_m(x)\,\upsilon_m.
\eqn\bunhou
$$
Since the four eigenvectors are independent, the equation
(\bunhou) must hold for each $m$, and consequently we obtain
$g_m(x) = e^{\mp x/(\lambda a_m)}$ or
$$
\Psi^{(m)}_{\pm i}(x) = e^{\mp x/(\lambda a_m)} \upsilon_m, 
\qquad  m = 1,\ldots, 4.
\eqn\eifun
$$
In view of (\fmgm), we find 
$\det(\Gamma-aI) = (a+1)^2(a-1)^2$ and hence the eigenvalues of $\Gamma$ are
$\pm 1$ (both doubly degenerate).  Accordingly, 
the deficiency indices are found to be $(4, 4)$, implying that
the supercharge $Q$ admits a $U(4)$ parameter family of
self-adjoint domains for systems of 
two intervals. 

The appearance of the $U(4)$ family may be understood directly from
the condition (\marimo) which reads
$$
\Psi^\dagger(l)\Gamma\Psi(l) - 
\Psi^\dagger(0)\Gamma\Psi(0)= 0.
\eqn\yago
$$
Exploiting the freedom of conjugation by 
$\Sigma$, $S$ and $V$ (and also
${\cal F}^\pm_i$, if necessary), we may take $A={I_2}$ in $\Gamma$ with no
loss of generality. 
Then we see by an argument similar to reach (\bouzu) 
that (\yago) is ensured 
if the state satisfies the boundary condition at the ends of the intervals,
$$
(U-I)\Psi_- +c_0(U+I)\Psi_+ =0, \qquad U \in U(4),
\eqn\qsabc
$$
where we have introduced $\Psi_-=(\Psi_1(0), -\Psi_1(l))^T$,  
$\Psi_+=(\Psi_2(0),
\Psi_2(l))^T$ and a dimensionless real constant
$c_0 \ne 0$.  The choice of the boundary condition is indeed
specified by the matrix $U$ belonging to $U(4)$.

{}For systems of two lines, on the other hand, the condition for $Q$ to
be self-adjoint arises only from the contribution at $x = 0$ in the foregoing
argument.  Thus the deficiency
indices of the operator $Q$ become $(2, 2)$ (since half of the
eigenfunctions (\eifun) are no longer square integrable), and hence one obtains
a $U(2)$
family of the self-adjoint domains for the supercharge $Q$.  These are
characterized by the boundary condition (\qsabc) restricted to the half 
subspace associated with
$x = 0$, where the group for the matrix $U$ reduces to
$U(2)$. 

In seeking a SUSY system in the preceding sections, we have tacitly assumed that the
supercharge is defined, at least, on eigenstates of some, self-adjoint Hamiltonian. 
In regard to this, it is assuring and also interesting to observe that, in fact, the
domain of any self-adjoint Hamiltonian is contained in
some domain of a self-adjoint supercharge.   To see this, for brevity 
we consider
only the conditions associated with the endpoint
$x = 0$ for a general characteristic matrix $U \in U(4)$, since an analogous
argument applied to the other endpoint
$x = l$.  

Now, given a $U$, we choose some orthonormal set of eigenvectors
$f_i$, $i = 1, \ldots, 4$, of $U$, that is, $Uf_i =e^{i\theta_i}f_i$
with the vectorial inner-product $\langle f_i, f_j\rangle = \delta_{ij}$.
In terms of these, we consider the decomposition of the boundary vectors 
$\Psi(0)= \sum_{i = 1}^{4}{\langle f_i, \Psi(0)\rangle f_i}$ and
$\Psi^\prime(0)= \sum_{i = 1}^{4}{\langle f_i, \Psi^\prime(0)\rangle f_i}$.
The boundary condition (\bouzu) then becomes
$$
\sum_{i=1}^{4}{\bigl[ (e^{i\theta_i}-1)\langle f_i, \Psi(0)\rangle
+ iL_0(e^{i\theta_i}+1)\langle f_i, \Psi^\prime(0)\rangle \bigr] f_i } = 0.
\eqn\mizusumasi
$$
Since the eigenvectors are independent each other, the
condition (\mizusumasi) must hold for each eigenvector, separately.   But since 
a SUSY system has $-1$ for two 
eigenvalues, which we choose $e^{i\theta_3}$ and $e^{i\theta_4}$, we obtain
$$
\langle f_i, \Psi(0)\rangle  = 0, 
\qquad
\hbox{for} \quad  i = 3, \, 4.
\eqn\yamori
$$
To get a condition for other eigenvectors, $f_1$ and $f_2$, 
we replace $D$ with the general $U$ in
(\hamati) (note that the diagonalization exploited in section 2 may not 
be available in
the interval systems) and multiply (\hamati) by
$f_i$ from the right, and similarly (\hamati) by
$f_j^\dagger U^\dagger$ from the left to find
$$
f_j^\dagger\Gamma f_i = 0,
\qquad
\hbox{for} \quad  i, j = 1, \, 2.
\eqn\tanisi
$$
We then see that
the second term in (\yago), decomposed similarly 
in terms of the eigenvectors, becomes
$$
\Psi^\dagger(0)\Gamma\Psi(0) =
\sum_{i,j = 1}^{4}{\langle f_i, \Psi(0)\rangle \langle \Psi(0), f_j\rangle\,
f_j^\dagger\Gamma f_i} = 0.
\eqn\yadokari
$$
on account of the identities (\yamori) and (\tanisi).  By a similar argument, the
first term in (\yago) can also be seen to vanish.

It is important to recognize that (\tanisi) provides a relation between the 
characteristic matrix $U$ and the corresponding supercharge $Q$, whereas 
(\yamori) furnishes the self-adjoint boundary condition for the
supersymmetric Hamiltonian
$H$.  (For intervals we need to add extra conditions at $x = l$ analogously.)   
Thus,
what we have seen here is that the combination of the two conditions are sufficient
to ensure that the supercharge
$Q$ be self-adjoint, namely, the domain
of such $H$ is contained in the domain of a self-adjoint $Q$.

\ve
\secno=5 \meqno=1

\centerline{\bf 5. Conclusion and discussions}
\medskip

In the present paper we have studied the possibility of SUSY 
in systems consisting of a pair of
lines/intervals each of which possesses a point singularity.  These two
point singularities are in general different and can be specified by the matrix 
$U \in U(2) \times U(2)$ given in
(\bloblo).  The line systems are thus specified by such $U$ and found to possess
an $N = 1$ SUSY if the matrix 
$U$, when properly diagonalized into $\bar D$, takes the form (\dbloblo).  The
SUSY is enhanced to
$N = 4$  if $B = 0$, {\it i.e.}, 
if the two angle parameters $\theta_1$ and $\theta_2$ in
$\bar D$ and the constant $\mu$ in the supercharge $Q$ satisfy
(\hondawake).   The SUSY is 
broken except for a restricted class of point singularities
allowing for supercharges with $B = 0$.   To specify the
interval systems, besides the matrix $U$ we further need 
an extra matrix
$D_l \in [U(1)]^4$ that characterizes the walls at the two ends.  
Exploiting the
freedoms in the two matrices, we have found various types of SUSY
systems which have either (good or broken) $N = 2$ or $N = 4$ SUSY. 
These newly found SUSY systems include the known SUSY system with the Dirac
$\delta(x)$-potential as a special case of type (B1) for line systems and also provide
a similar example for interval systems as type (c3).  The 
spectra and the SUSY properties for all of these SUSY systems are summarized in the
table in Appendix B.

One of the important points in our analysis is that our
supercharge $Q$ in (\supsup) has the $\mu$-term in addition to the
conventional derivative term.  This
$\mu$-term allows 
us to acquire the variety of the SUSY systems having two
independent scale parameters $L(\theta_1)$ and $L(\theta_2)$, 
at the expense of the
constant shift of energy by
$\mu^2$ in the Hamiltonian for realizing the standard SUSY algebra (\ancc).  
Without the $\mu$-term, we obtain only a restricted class of SUSY systems 
given by a combination of Dirichlet and Neumann boundary conditions 
[\Kobe].  For interval systems, we have essentially exhausted the SUSY systems
for the $B = 0$ case, but a large number
of novel SUSY systems for the general $B \ne 0$ case will exist under
the $\mu$-term.  The introduction of the $\mu$-term may in a sense be regarded as a
generalization of  the SUSY potential in the Witten model.  In fact, our supercharge
$Q$ in (\izanagi) has a structure analogous to (the doubly graded form of) 
the one used in 
the Witten model,  Moreover, for $B = 0$ the
supercharge $Q$ reduces essentially to the supercharge
of the Witten model with a constant SUSY potential [\Kobe].  

Another point to be noted is the notion of SUSY itself.  Namely, to seek SUSY
systems we adopted the criterion [\TUIT] that eigenstates of the
Hamiltonian $H$ satisfy the same boundary condition even after the 
SUSY transformation by $Q$ is performed.  This is a necessary condition
for the complete SUSY invariance of the boundary condition ({\it i.e.}, valid for
any state, not just for energy eigenstates) but may be shown to be sufficient,
too.
We note that the issues such as the complete SUSY invariance
or the self-adjointness of
the supercharge $Q$ discussed in section 4 have not been fully addressed 
in generic SUSY quantum
mechanics, despite that they become important 
if we wish to put SYSY  
on a firm basis in systems with boundaries or singularities.  The relation
between the self-adjoint domains of $Q$ and $H$ pointed out in this paper may
provide a first step toward the full investigation on these issues.

Related to the above two points, we mention that independent
supercharges fulfilling the criterion may not, in general, admit a basis set
realizing the orthogonal SUSY algebra (\ototachibana).  In our analysis, we have
defined the number $N$ of SUSY by the number of supercharges fulfilling the
orthogonal SUSY algebra, rather than by the total number of the independent
supercharges.  One may instead accept the full SUSY algebra --- though it may be
fairly involved --- formed by the entire set of the supercharges as an extended
version of SUSY, adopting the total number of the charges for the number
$N$ of SUSY.  

To extend our work, perhaps the most
straightforward is to put point singularities on more complicated one dimensional 
systems,
such as a circle or networks with loops and vertices.  We may also add a potential
$V(x)$ to our systems without changing our argument, as long as $V(x)$ is regular
at the singularities.  In fact, the possibility of SUSY of a circle system with two point
singularities has been studied recently in Ref.[\Kobe], where the introduction of
regular potentials in the framework of the Witten model has also been discussed.  
We wish to stress, however, that the regularity of the potential is not essential,
that is, even if
$V(x)$ is singular (like the Coulomb
potential), we can treat it by generalizing slightly our
procedure of assigning the boundary/connection conditions at the singularities
[\TCF].  We believe that various novel SUSY systems will be
obtained under singular potentials if one employs the generalized approach developed in
this paper.

\bigskip
\noindent
{\bf Acknowledgement:}
I.T.~thanks T. F\"{u}l\"{o}p for useful discussions.
This work has been supported in part by
the Grant-in-Aid for Scientific 
Research on Priority Areas (No.~13135206) by
the Japanese
Ministry of
Education, Science, Sports and Culture.

\ve
\centerline{{\bf Appendix A. Solutions for (\gamga) and (\omom)}}
\medskip

\secno=0 \appno=1 \meqno=1 

\noindent{\bf A.1. type (a)}
\medskip

The choice of (a) for $D$ and $D_l$ implies $S = S_l = I$, and hence (\gamga) is
fulfilled if
$$
V_1^\dagger A=A_l.
\eqn\gaa
$$
On the other hand, from (\omom) we have
$V_1^\dagger{K} A={K}_l A_l$ which, in view of (\gaa), becomes
$(V_1^\dagger {K} V_1 - {K}_l)A_l=0$ or  
$$
V_1^\dagger {K} V_1 = {K}_l,
\eqn\gob
$$
because $A_l\in U(2)$ is invertible.  With ${K}$ and ${K}_l$
in (\lamdef), 
(\gob) is seen to be
$$
\left(1+s_2-s_1^l-s_2^l\right){I_2}+
\left(1-s_2\right)V_1^\dagger\sigma_3V_1 - \left(s_1^l-s_2^l\right)\sigma_3 = 0,
\eqn\wiski
$$
that is,
$$
s_2-s_1^l-s_2^l+1=0, \qquad 
\left(1-s_2\right)V_1^\dagger \sigma_3V_1 = \left(s_1^l-s_2^l\right)\sigma_3.
\eqn\wiwiwi
$$
Using the parametrization of $V_1$ in (\vpara), the second equation of (\wiwiwi)
becomes
$$
\left(1-s_2\right)e^{-i\delta_1\sigma_2} 
\sigma_3 e^{i\delta_1\sigma_2} = \left(s_1^l-s_2^l\right)\sigma_3,
\eqn\goc
$$
which is satisfied under the three cases:
$$
\eqalign{
&(1)\quad s_2=1, \quad s_1^l=s_2^l, \qquad
(2)\quad s_2+s_1^l-s_2^l=1, \quad \delta_1=0, \quad (s_1^l-s_2^l\neq 0), \cr
&(3)\quad s_2-s_1^l+s_2^l=1,  \quad
\delta_1=\pi/2 , \quad(s_1^l-s_2^l\neq 0).
}
\eqn\no
$$
Combining these with (\gaa) and the fist equation of (\wiwiwi), we obtain the
solutions (a1) -- (a3).

\bigskip
\noindent{\bf A.2. type (b)}
\medskip

In this case we have $S = X$ and $S_l = I$.  Then (\gamga) is solved by
$$
V_2^\dagger A=A_l^\dagger.
\eqn\gba
$$
Similarly to type (a), with the solution (\gba) the condition (\omom) implies
$A^\dagger {K} A=-{K}_l$
from which we find
$$
s_2+s_1^l+s_2^l+1=0, \qquad
\left(1-s_2\right)A^\dagger\sigma_3 A+\left(s_1^l-s_2^l\right)\sigma_3=0.
\eqn\gobc
$$
Using (\vpara), the second equation of (\gobc) becomes
$$
\left(1-s_2\right)e^{-i\beta\sigma_2}\sigma_3 e^{i\beta\sigma_2}+
\left(s_1^l-s_2^l\right)\sigma_3=0,
\eqn\gobd
$$
which is satisfied under the three cases:
$$
\eqalign{
&(1)\quad s_2=1, \quad s_1^l=s_2^l, \qquad 
 (2)\quad s_2=s_1^l=-1, \quad s_2^l=1, \quad \beta=0 \cr
&(3)\quad s_2=s_2^l=-1, \quad s_1^l=1, \quad \beta=\pi/2.
}
\eqn\no
$$
Combining these with (\gba) and the first equation of (\gobc), we obtain 
(b1) -- (b3).

\bigskip
\noindent{\bf A.3. type (c)}
\medskip

Here we have $S = Y$ and $S_l = I$.
Then (\gamga) implies
$$
\eqalign{
&\sigma_+A\sigma_2\sigma_--\sigma_-\sigma_2A^\dagger\sigma_+ =
\sigma_+\sigma_2A\sigma_--\sigma_-A^\dagger\sigma_2\sigma_+=0, \cr
&V_1^\dagger\left(\sigma_+A\sigma_-
-\sigma_-\sigma_2A^\dagger\sigma_2\sigma_+\right)V_2=A_l, 
}
\eqn\acd
$$
where $\sigma_\pm$ are defined in (\yamatotakeru), 
whereas (\omom) gives
$$
\eqalign{
&\sigma_+A\sigma_2\sigma_-+\sigma_-\sigma_2A^\dagger\sigma_+=
\sigma_+\sigma_2A\sigma_-+\sigma_-A^\dagger\sigma_2\sigma_+=0, \cr
&V^\dagger_1\left(\sigma_+A\sigma_-
+s_2\sigma_-\sigma_2A^\dagger\sigma_2\sigma_+\right)V_2=\left(s_1^l\sigma_+
+s_2^l\sigma_-\right)A_l.
}
\eqn\bcd
$$
The first equations of (\acd) and (\bcd) are satisfied by
$
\beta=\pi/2.
$
{}From the second equations of (\acd) and (\bcd), we have
$$
\left((1-s_1^l)\sigma_++(1-s_2^l)\sigma_-\right)V_1^\dagger\sigma_+A\sigma_- +\left((s_2+s_1^l)\sigma_++(s_2+s_2^l)\sigma_- \right)V_1^\dagger\sigma_-\sigma_2A^\dagger\sigma_2\sigma_+=0,
\eqn\dcd
$$
which implies
$$
\eqalign{
&(1-s_1^l)\sigma_+V_1^\dagger\sigma_+A\sigma_-=
(1-s_2^l)\sigma_-V_1^\dagger\sigma_+A\sigma_- = 0, \cr
&(s_2+s_1^l)\sigma_+V_1^\dagger\sigma_-\sigma_2A^\dagger\sigma_2\sigma_+ 
= (s_2+s_2^l)\sigma_-V_1^\dagger\sigma_-\sigma_2A^\dagger\sigma_2\sigma_+=0,
}
\eqn\ecd
$$
form which we obtain
$$
\eqalign{
(1-s_1^l)\sin\beta\cos\delta_1=(1-s_2^l)\sin\beta\sin\delta_1&=0, \cr
(s_2+s_1^l)\sin\beta\sin\delta_1=(s_2+s_2^l)\sin\beta\cos\delta_1&=0.
}
\eqn\fcd
$$
By $\beta=\pi/2$, eq.(\fcd) is seen to be satisfied under the there cases:
$$
\eqalign{ 
&(1)\quad\delta_1=0, \quad s_2 = s_1^l = 1, \quad
s_2^l=-1, \quad
(2)\quad\delta_1=\pi/2, \quad s_2=s_2^l=1, \quad s_1^l=-1, \cr
&(3)\quad s_2=-1, \quad s_1^l=s_2^l=1.
}
\eqn\ccse
$$
Solving the second equation of (\acd) for the three cases of (\ccse), we obtain (c1) -- (c3).

\bigskip
\noindent{\bf A.4. type (d)}
\medskip

Here we have $S = S_l = Y$, and 
from (\gamga) we find
$$
\eqalign{
&(\sigma_+V_1^\dagger\sigma_-+\sigma_-V_2^T\sigma_+)\sigma_2A^\dagger(\sigma_+V_1\sigma_++\sigma_-V_2^*\sigma_-) \cr
&\quad\,\,
-(\sigma_+V_1^\dagger\sigma_++\sigma_-V_2^T\sigma_-)A\sigma_2(\sigma_-V_1\sigma_++\sigma_+V_2^*\sigma_-)=0,
\cr
&(\sigma_+V_1^T\sigma_++\sigma_-V_2^\dagger\sigma_-)A^\dagger\sigma_2(\sigma_-V_1^*\sigma_++\sigma_+V_2\sigma_-)
\cr
&\quad\,\,
-(\sigma_+V_1^T\sigma_-+\sigma_-V_2^\dagger\sigma_+)\sigma_2A(\sigma_+V_1^*\sigma_++\sigma_-V_2\sigma_-)=0,
\cr
&(\sigma_+V_1^\dagger\sigma_++\sigma_-V_2^T\sigma_-)A(\sigma_+V_1^*\sigma_++\sigma_-V_2\sigma_-)
\cr
&-(\sigma_+V_1^\dagger\sigma_-+\sigma_-V_2^T\sigma_+)\sigma_2A^\dagger\sigma_2(\sigma_-V_1^*\sigma_++\sigma_+V_2\sigma_-)=A_l.
}
\eqn\naninani
$$
{}From (\omom) we have 
$$
\eqalign{
&(\sigma_+V_1^\dagger\sigma_-+\sigma_-V_2^T\sigma_+)\sigma_2A^\dagger(\sigma_+V_1\sigma_++s_2\sigma_-V_2^*\sigma_-) \cr
&\quad\,\,+(\sigma_+V_1^\dagger\sigma_++s_2\sigma_-V_2^T\sigma_-)A\sigma_2(\sigma_-V_1\sigma_++\sigma_+V_2^*\sigma_-)=0,
\cr
&(\sigma_+V_1^T\sigma_++\sigma_-V_2^\dagger\sigma_-)A^\dagger\sigma_2(\sigma_-V_1^*\sigma_++s_2\sigma_+V_2\sigma_-)
\cr
&\quad\,\,+(\sigma_+V_1^T\sigma_-+s_2\sigma_-V_2^\dagger\sigma_+)\sigma_2A(\sigma_+V_1^*\sigma_++\sigma_-V_2\sigma_-)=0,
\cr
&(\sigma_+V_1^\dagger\sigma_++s_2\sigma_-V_2^T\sigma_-)A(\sigma_+V_1^*\sigma_++\sigma_-V_2\sigma_-)
\cr
&+(\sigma_+V_1^\dagger\sigma_-+\sigma_-V_2^T\sigma_+)\sigma_2A^\dagger\sigma_2(\sigma_-V_1^*\sigma_++s_2\sigma_+V_2\sigma_-)=(s_1^l\sigma_++s_2^l\sigma_-)A_l.
}
\eqn\nanio
$$
{}From the first equation of (\naninani) we obtain
$$
\eqalign{
\sigma_+V_1^\dagger(\sigma_-\sigma_2A^\dagger\sigma_+-\sigma_+A\sigma_2\sigma_-)V_1\sigma_+&=0,
\cr
\sigma_+V_1^\dagger(\sigma_-\sigma_2A^\dagger\sigma_--\sigma_+A\sigma_2\sigma_+)V_2^*\sigma_-&=0,
\cr
\sigma_-V_2^T(\sigma_+\sigma_2A^\dagger\sigma_--\sigma_-A\sigma_2\sigma_+)V_2^*\sigma_-&=0,
}
\eqn\dema
$$
whereas from the first equation of (\nanio) we find
$$
\eqalign{
\sigma_+V_1^\dagger(\sigma_-\sigma_2A^\dagger\sigma_++\sigma_+A\sigma_2\sigma_-)V_1\sigma_+&=0,
\cr
\sigma_+V_1^\dagger(s_2\sigma_-\sigma_2A^\dagger\sigma_-
+\sigma_+A\sigma_2\sigma_+)V_2^*\sigma_-&=0,
\cr
\sigma_-V_2^T(\sigma_+\sigma_2A^\dagger\sigma_-+\sigma_-A\sigma_2\sigma_+)V_2^*\sigma_-&=0. }
\eqn\demb
$$
The first and third equations of (\dema) and (\demb) imply
$$
\eqalign{
\sigma_+V_1^\dagger\sigma_-\sigma_2A^\dagger\sigma_+V_1\sigma_+
=\sigma_+V_1^\dagger\sigma_+A\sigma_2\sigma_-V_1\sigma_+&=0, \cr
\sigma_-V_2^T\sigma_+\sigma_2A^\dagger\sigma_-V_2^*\sigma_-
=\sigma_-V_2^T\sigma_-A\sigma_2\sigma_+V_2^*\sigma_-&=0.
}
\eqn\demc
$$
(\demc) is satisfied by
$$
\sin2\delta_1\cos\frac{\beta}{2}=0, \qquad \sin2\delta_2\cos\frac{\beta}{2}=0.
\eqn\odema
$$
{}From the second equations of (\dema) and (\demb), we have
$$
\eqalign{
\sin\delta_1\cos\delta_2\sin\frac{\beta}{2}=\cos\delta_1\sin\delta_2\sin\frac{\beta}{2}=0 \quad &{\rm for} \enspace s_2=1, \cr
(\tan\delta_1e^{i\xi}-\tan\delta_2e^{-i\xi})\sin\frac{\beta}{2}=0 \quad &{\rm for} \enspace s_2=-1.
}
\eqn\odemb
$$
The second equations of (\naninani) and (\nanio) hold automatically 
by (\odema) and (\odemb) which are satisfied under the six cases:
$$
\eqalign{
&(1)\quad \delta_1=\delta_2=0, \qquad
(2)\quad \delta_1=\delta_2=\pi/2, \qquad
(3)\quad \delta_1=0, \quad \delta_2=\pi/2, \quad \beta=0, \cr
&(4)\quad \delta_1=\pi/2, \quad \delta_2=0, \quad \beta=0, \cr
&(5)\quad \delta_1=\delta_2\neq0\enspace{\rm or}\enspace\pi/2, \quad \beta=\pi, \quad \xi=0
\enspace {\rm or} \enspace \pi \quad {\rm for} \enspace s_2=-1,\cr
&(6)\quad \delta_2=\pi-\delta_1\neq\pi/2, \quad \beta=\pi, \quad \xi=\pi/2 \enspace {\rm or}
\enspace 3\pi/2 \quad {\rm for} \enspace s_2=-1. }
\eqn\odemc
$$
Solving the third equations of (\naninani) and (\nanio) for 
the six cases of (\odemc), we obtain (d1) -- (d6).

\ve
\centerline{{\bf Appendix B. Spectral and SUSY properties
}}
\medskip

The following table summarizes the spectral as well as SUSY properties of the
various types of SUSY systems obtained for two lines and two intervals.

\medskip 
\noindent
$\bullet$ Two lines
\smallskip
\halign{ #\hfil& \quad #\hfil& \quad #\hfil& \quad #\hfil& \quad #\hfil&\quad #\hfil \cr
\noalign{\smallskip\hrule\smallskip}
\hfil Type&\hfil Number of&\hfil SUSY&\hfil Spectrum of 
&\hfil Degeneracy of &\hfil Number of
\cr & \hfil supercharges& \hfil &\hfil regular series&\hfil regular series&\hfil
isolated  eigenstates \cr
\noalign{\smallskip\hrule\smallskip}
\hfil(A1)& \hfil $N$ = 1&\hfil $\times$&
\hfil$k>0$& \hfil 4&\hfil 0 or 1 or 2 \cr
\hfil(A2)& \hfil $N$ = 1&\hfil $\times$&
\hfil$k>0$& \hfil 4&\hfil 0 or 2 \cr
\hfil(B1)& \hfil $N$ = 4&\hfil $\bigcirc$&
\hfil$k>0$& \hfil 4&\hfil 1 or 2 \cr
\hfil(B2)& \hfil $N$ = 4& \hfil $\times$&
\hfil $k>0$& \hfil 4&\hfil 0 \cr
\noalign{\smallskip\hrule\smallskip}
}

\medskip
\noindent
$\bullet$ Two intervals
\smallskip
\halign{ #\hfil& \quad #\hfil& \quad #\hfil& \quad #\hfil& \quad #\hfil& \quad #\hfil \cr
\noalign{\smallskip\hrule\smallskip}
\hfil Type&\hfil Number of&\hfil SUSY&\hfil Spectrum of&\hfil Degeneracy of
&\hfil Number of
\cr & \hfil supercharges& \hfil &\hfil regular series&\hfil regular series&\hfil isolated  
eigenstates \cr
\noalign{\smallskip\hrule\smallskip}
\hfil(a1)& \hfil $N$ = 4&\hfil $\bigcirc$&
\hfil$k_n$& \hfil 4&\hfil 2 \cr
\hfil(a2)& \hfil $N$ = 4&\hfil $\bigcirc$&
\hfil$k_n$& \hfil 4&\hfil 2 \cr
\hfil(a3)& \hfil $N$ = 4&\hfil $\bigcirc$&
\hfil$k_n$& \hfil 4&\hfil 2 \cr
\hfil(b1)& \hfil $N$ = 2&\hfil $\times$&
\hfil$k_n^+$, $k_n^-$& \hfil 2+2&\hfil 0 or 2 \cr
\hfil(b2)& \hfil $N$ = 2&\hfil $\times$&
\hfil$k_n^+$, $k_n^-$& \hfil 2+2&\hfil 0 or 2 \cr
\hfil(b3)& \hfil $N$ = 4&\hfil $\times$&
\hfil$k_n^-$& \hfil 4&\hfil 0 or 4 \cr
\hfil(c1)& \hfil $N$ = 2&\hfil $\bigcirc$&
\hfil$k_n$, $k_n^-$& \hfil 2+2&\hfil 1 or 3 \cr
\hfil(c2)& \hfil $N$ = 2&\hfil $\bigcirc$&
\hfil$k_n$, $k_n^-$& \hfil 2+2&\hfil 1 or 3 \cr
\hfil(c3)& \hfil $N$ = 2&\hfil $\bigcirc$&
\hfil$k_n$, $k_n^+$& \hfil 2+2&\hfil 1 or 3 \cr
\hfil(d1)& \hfil $N$ = 2&\hfil $\bigcirc$&
\hfil$k_n$, $k_n^{-s_2}$& \hfil 2+2&\hfil 1 or 3 \cr
\hfil(d2)& \hfil $N$ = 2&\hfil $\bigcirc$&
\hfil$k_n$, $k_n^-$& \hfil 2+2&\hfil 1 or 3 \cr
\hfil(d3)'& \hfil $N$ = 2&\hfil $\times$&
\hfil $k_n +\delta_1/l$& \hfil 2&\hfil 0 \cr
\hfil(d4)'& \hfil $N$ = 2&\hfil $\times$&
\hfil $k_n +\delta_1/l$& \hfil 2&\hfil 0 \cr
\hfil(d5)& \hfil $N$ = 4&\hfil $\bigcirc$& \hfil $k_n$& \hfil 4&\hfil 2 \cr
\hfil(d6)& \hfil $N$ = 4&\hfil $\times$&
\hfil $k_n^-$& \hfil 4&\hfil 0 or 4 \cr
\noalign{\smallskip\hrule\smallskip}
}
\medskip
\noindent
Note: \lq 2+2\rq{} means that two distinct types of doubly degenerate
eigenstates exist;
$\bigcirc$ and $\times$ 
denote good and broken SUSY, respectively; $k_n$ and $k^\pm_n$ are given in
(\kzero) and (\kpm); and (d3)' and (d4)' refer to
the special $\theta = 0$ case in type (d3) and (d4), respectively.

\baselineskip= 15.5pt plus 1pt minus 1pt
\parskip=5pt plus 1pt minus 1pt
\tolerance 8000
\vfill\eject\immediate\closeout\reffile
\centerline{{\bf References}}\bigskip\frenchspacing%
\input refs.tmp\vfill\eject\nonfrenchspacing

\bye